\documentclass[dvips,10.5pt]{imsart}

\RequirePackage[OT1]{fontenc}
\RequirePackage{amsthm,amsmath,natbib,amssymb}
\RequirePackage{graphicx}
\RequirePackage[colorlinks]{hyperref}
\usepackage{a4wide}
\startlocaldefs
\numberwithin{equation}{section}
\theoremstyle{plain}

\endlocaldefs

\newcommand{\te}{\theta}
\newcommand{\Te}{\Theta}
\begin{document}

\begin{frontmatter}
\title{The assessment and planning of non-inferiority trials for retention of effect hypotheses - towards a general approach}
\runtitle{Evaluating and planning the RET}
%\thankstext{T1}{Footnote to the title with the `thankstext' command.}

\begin{aug}
\author{\fnms{M.} \snm{Mielke}\thanksref{t1}\ead[label=e1]{mmmielke@math.uni-goettingen.de}}
\and
\author{\fnms{A.} \snm{Munk}\ead[label=e2]{munk@math.uni-goettingen.de}}

\address{Institute for Mathematical Stochastics, University of G\"ottingen}
%\printead{e1}}

\thankstext{t1}{\textbf{Corresponding Author}: Matthias Mielke, University of G\"ottingen, \textit{mmielke@math.uni-goettingen.de}}
\runauthor{Mielke \& Munk }

\affiliation{Some University and Another University}

\end{aug}

\begin{abstract}
The objective of this paper is to develop statistical methodology for planning and evaluating three-armed non-inferiority trials for general retention of effect hypotheses, where the endpoint of interest may follow any (regular) parametric distribution family. This generalizes and unifies specific results for binary, normally and exponentially distributed endpoints. We propose a Wald-type test procedure for the retention of effect hypothesis (RET), which assures that the test treatment maintains at least a proportion $\Delta$ of reference treatment effect compared to placebo. At this, we distinguish the cases where the variance of the test statistic is estimated unrestrictedly and restrictedly to the null hypothesis, to improve accuracy of the nominal level. We present a general valid sample size allocation rule to achieve optimal power and sample size formulas, which significantly improve existing ones. Moreover, we propose a general applicable rule of thumb for sample allocation and give conditions where this rule is theoretically justified. The presented methodologies are discussed in detail for binary and for Poisson distributed endpoints by means of two clinical trials in the treatment of depression and in the treatment of epilepsy, respectively. $R$-software for implementation of the proposed tests and for sample size planning accompanies this paper. 
\end{abstract}

%\begin{keyword}[class=AMS]
%\kwd[Primary ]{}
%\kwd[; secondary ]{}
%\end{keyword}

%\tableofcontents
\end{frontmatter}

\section{Introduction}\label{sec:intro}

The aim of a non-inferiority trial is to demonstrate that the efficacy of a test treatment relative to a reference one does not fall below a clinically relevant value. For selective fundamental references we refer to Jones et al. (1996), R\"ohmel (1998), D'Agostino (2003) and Munk \& Trampisch (2005). In this work we focus on the direct comparison of a test and reference group. To this end, the inclusion of a concurrent placebo group is recommended if there are no ethical concerns, i.e. the patients are not harmed by deferral of therapy and are fully informed about alternative (see e.g. Temple \& Ellenberg (2000) and Hypericum Depression Trial Study Group (2004)), to ensure for assay sensitivity of the trial. Such a design, including a (T)est, (R)eference and (P)lacebo group, has been coined by Koch \& R\"ohmel (2004) as gold standard design. 

\smallskip
\noindent\textbf{Retention of effect hypothesis:} To demonstrate non-inferiority in the gold standard design we consider the retention of effect type hypothesis
\begin{eqnarray}
&H_{0}:\;\te_T-\te_P\;\leq \;\Delta \cdot (\te_R-\te_P) \nonumber\\
{\rm vs.}& \label{eqn:intro_hyp} \\
& H_{1}:\;\te_T-\te_P\;> \;\Delta \cdot (\te_R-\te_P), \nonumber
\end{eqnarray}
where $\theta_k\in\Theta\subseteq \mathbb{R}$, $k=T,R,P$, is the parameter of interest, representing the efficacy of a treatment, and $\Delta\in[0,\infty)$ a fixed constant expressing the amount of the active control effect relative to placebo, which should be retained. For a discussion of various issues encountered with the choice of $\Delta$ we refer to Lange \& Freitag (2005) and the references given there, who provide a systematic review of 332 published non-inferiority studies. Examples for $\theta_k$ are (a) $\theta_k=\pi_k$ the success probability of a binary endpoint representing for example if the patient achieves remission (Kieser \& Friede, 2007), (b) $\theta_k=\lambda_k$ the expectation of an exponentially distributed endpoint representing for example the time until healing or remission (Mielke et al., 2008), (c) $\theta_k=\mu_k$ the expectation of a normally distributed endpoint representing for example the FCV (forced vital capacity) in a trial on mildly asthmatic patients (Pigeot et al., 2003).  Note, that in this set up we presume that large values of $\theta_k$ are associated with higher efficacy of the treatment. Compared to absolute hypotheses, e.g. $H_0: \theta_T \leq \theta_R - \Delta$ with $\Delta>0$, the advantage of the hypothesis (\ref{eqn:intro_hyp}) is that it is invariant with respect to rescaling or shifts of the parameters $\theta_k$, i.e. the margin $\Delta$ must not be readjusted to the changes of parametrization. Thus, the margin $\Delta$ is standardized in that sense and therewith it could easily be compared for different hypothesis and applications, respectively. Further, it has an intuitive and clear interpretation. Rejecting $H_{0}$ implies to claim that the test treatment achieves at least $\Delta\cdot 100\%$ of the active control effect, at which both are compared relatively to placebo. Rewriting the alternative in (\ref{eqn:intro_hyp}) as 
\begin{eqnarray*}
& H_{1}:\;\te_T\;> \;\Delta \cdot \te_R + (1-\Delta)\cdot \te_P 
\end{eqnarray*}
illustrates that in this case the test treatment effect is greater than a convex combination of the reference and the placebo effect if $0\leq \Delta \leq 1$. This includes two extremal cases: For $\Delta=1$ we obtain superiority of the test treatment to the reference one (at least $\Delta=100\%$ of the reference effect is retained) and for $\Delta=0$ superiority of the test treatment to placebo. 

As mentioned above for binary endpoints a typical choice is $\theta_k=\pi_k$, the success probability. However, in practical application also transformations of the success probability are of interest, e.g. $\log(\pi_k),\: \pi_k/(1-\pi_k)$, $\log(\pi_k/(1-\pi_k))$ or just $-\pi_k$ in case of a mortality rate.  For a comprehensive discussion of several hypotheses for binary endpoints see R\"ohmel \& Mansmann (1999). In order to formalize this we modify the hypothesis (\ref{eqn:intro_hyp}) to  
\begin{eqnarray}
&H_{0,h(\theta_k)}:\; h(\te_T)-h(\te_P)\;\;\leq \;\;\Delta \cdot \left(h(\te_R)-h(\te_P)\right)\nonumber\\
{\rm vs.}& \label{eqn:g_hyp}\\
&H_{1,h(\theta_k)}:\; h(\te_T)-h(\te_P)\;\;> \;\;\Delta \cdot \left(h(\te_R)-h(\te_P)\right)\nonumber 
\end{eqnarray}
where $\theta_k\in\Theta\subseteq \mathbb{R}^d$, $k=T,R,P$, determines the distribution of our endpoints of interest. Here, $h(\cdot)$ is a differentiable, strictly monotone, real-valued function on the parameter space $\Theta$ measuring the efficiency of a treatment whereas larger values of $h(\cdot)$ correspond to higher efficiency. In the following, we will omit the alternatives and only state the null hypotheses.

\smallskip
\noindent\textbf{Aim and scope:} The aim of this work is to provide a general testing methodology based on Wald's maximum likelihood asymptotic to the general retention of effect hypotheses (\ref{eqn:g_hyp}). This, among others, includes the above mentioned situations as special cases. In addition, we obtain tests for Poisson distributed endpoints (for careful discussion see Section \ref{sec:ex_pois}). Moreover, we discuss the issue of sample size planning and we provide in large generality formulas for optimal allocation of samples and accurate approximations for the determination of sample sizes in order to guarantee a certain power. We show that this requires the computation of Kullback-Leibler divergence minimizer in the null hypothesis to an alternative model.

\smallskip
\noindent\textbf{Complete test procedure:} To ensure assay sensitivity of the test procedure the hypothesis (\ref{eqn:intro_hyp}) is typically embedded in a complete test procedure, where in a first step a pretest for superiority of either the reference or the test treatment to placebo is performed, and in a second step the non-inferiority is investigated via (\ref{eqn:intro_hyp}). There is a vigorous discussion on which pretest is appropriate. For example Pigeot et al. (2003) carry out a pretest for superiority of the reference treatment to placebo, whereas Koch \& R\"ohmel (2004) perform the test for the test treatment to placebo, because the test treatment should not be blamed when the reference treatment could not beat placebo (Koch, 2005). 

It is important to note that it turns out as a common rule that the pretest is subordinated in the complete test procedure, in terms of that sample size planning can be performed via the non-inferiority test without adjustment to the pretest for superiority (see e.g. Mielke et al., 2008). This means the power of the non-inferiority test nearly coincides with the power of the complete test procedure for commonly used alternatives. In addition, the pretest represents a well-investigated  testing problem where the parameters of comparison coincide on the boundary of the hypothesis. Thus, we only focus in the following on the non-inferiority hypothesis (\ref{eqn:intro_hyp}) and keep the complete test procedure at the back of mind. 

\smallskip
\noindent\textbf{State of research:} Closely related to the retention of effect hypothesis (\ref{eqn:intro_hyp}) is the hypothesis where the treatment effect $\theta_T-\theta_R$ is evaluated relative to a historic active control effect $\tilde{\theta}_R-\tilde{\theta}_P$, which could not estimated concurrently, therefore. For a comprehensive discussion we refer to Holgrem (1999), Hauck \& Anderson (1999), Hasselblad \& Kong (2001), Rothmann et al. (2003) and Hung, Wang \& O'Neill (2009). The most problematic issue of such design is the necessity to project the active control effect in the current non-inferiority trial setting (Hung, Wang \& O'Neill, 2009). This issue is not present in the gold standard design, where the active control effect is estimated concurrently.  

A nonparametric version of the retention of effect hypothesis (\ref{eqn:intro_hyp}) was already considered by Koch \& Tangen (1999).  Pigeot et al. (2003) consider (\ref{eqn:intro_hyp}) for normally distributed endpoints. Subsequently, this type of hypothesis was discussed vigorously (see e.g. Hauschke \& Pigeot, 2005) and investigated for different types of endpoints. Koch \& R\"ohmel (2004) and Schwartz \& Denne (2006) also consider normally distributed endpoints and investigate (\ref{eqn:intro_hyp}) for $\theta_k$ equals the expectation $\mu_k$ of the groups $k=T,R,P$, respectively, under homogeneity of variance between the groups. Hasler et al. (2008) and Dette et al. (2009) extend these results to the case of heterogeneity of the group variances. Mielke et al. (2008) consider censored, exponentially distributed endpoints. Tang \& Tang (2004) and Kieser \& Friede (2007) investigate binary endpoints with $\theta_k$ equals the success probability $\pi_k$ of each group. In contrast to the normal and exponential case for binary endpoints sample size planning leaves open questions. In particular, the existing sample size formulas lack in precision, i.e. a deviation between exact and aspired power (cf. Kieser \& Friede, 2007). The additional difficulties for binary endpoints are mainly due to dependency of the variance on the parameters of interest, the success probabilities. In this work we will provide a general approach for general parametric models which allows to close this gap for binary endpoints as a special case.

\smallskip
\noindent\textbf{Content and organization:} This paper is organized as follows. In Section \ref{sec:example}, we discuss two clinical trials. First a trial in the treatment of depressions by investigating if the patients achieve remission at the treatment end (binary endpoints) measured by the Hamilton rating scale score of depression (HAM-D) and second a study in the treatment of epilepsy by investigating the number of seizures (Poisson distributed endpoints). In Section \ref{sec:test}, we present the general theory and derive a Wald-type test procedure for the generalized retention of effect hypothesis (\ref{eqn:g_hyp}), which we denote as \textit{Retention of Effect Wald-type Test (RET)} in the following. In Section \ref{sec:planning}, we derive sample size formulas and the (asymptotically) optimal allocation for planning a three-armed retention of effect trial. In particular, we include the important case where the variance is estimated restrictedly to the null hypothesis. This often improves the asymptotic approximation under the null hypothesis (see e.g. Farrington \& Manning (1990) and Tang, Tang \& Wang (2007)) and therefore is very popular in practice. For the presented sample size formulas we have determined the exact limit of the restricted ML-estimator, which has never been considered so far. As a major result this significantly improves the precision of the formulas, see exemplarily Table \ref{table_bin2}. The \textit{optimal allocation} when the variance is estimated unrestrictedly turns out to be
\begin{eqnarray}\label{eqn:intro:opt}
n_T^*:n_R^*:n_P^*\;=\;1:\Delta\:\frac{\sigma_{0,R}}{\sigma_{0,T}}:|1-\Delta|\:\frac{\sigma_{0,P}}{\sigma_{0,T}}\:,
\end{eqnarray}
where $\sigma_{0,k}$ is the variance within group $k$, $k=T,R,P$, under the alternative, specified later on in (\ref{sigma_0k}). Here, $n_k^*$ denotes the number of samples assigned to group $k=T,R,P$. This is shown to be valid in (essentially) any parametric family. Albeit the asymptotic power will change in general when the variance is estimated restrictedly, we argue that the \textit{optimal allocation} remains unchanged in an asymptotic sense even when the variance is estimated restrictedly to the null hypothesis. As the optimal allocation (\ref{eqn:intro:opt}) depends on the choice of the alternative we show in Section \ref{RuleOfThumb} that one may use the allocation $1:\Delta:(1-\Delta)$ as a very general rule of thumb, which is more appropriate in terms of power than the commonly used allocation 2:2:1 as well as the balanced allocation, if $\sigma^2_{0,P}/\sigma^2_{0,T}$ is (roughly) less than 2. It is important to note that this result is very general valid, independent of the distribution of the endpoints and of the formulation of the hypothesis (\ref{eqn:g_hyp}). 

In Section \ref{sec:examples_revisited}, we will revisit our examples introduced in Section \ref{sec:example} to demonstrate and to discuss the results of the previous sections in detail. We show that sample size reductions and therewith reductions in the costs of a trial with up to 20\% and more are possible by reallocating to the optimal allocation instead of a balanced or the commonly used 2:2:1 allocation. In particular, it turns out that our sample size formula for binary endpoints improves the precision of the existing one by Kieser \& Friede (2007) significantly in terms of that the exact power is close to the aspired one. In Section \ref{sec:software}, we briefly comment on $R$-software for analysis and planning of the RET, which we provide as supplementary material, in order to allow the reader to reproduce the presented results and to make the presented methodology directly applicable. Finally, we conclude with a discussion in Section \ref{sec:disc}.

\section{Examples}\label{sec:example}

In this section, we introduce two clinical non-inferiority trials, one in the treatment of epilepsy and the other one in the treatment of depression, and we define retention of effect hypotheses, which are of interest within these examples. 

\subsection{Binary endpoints: Treatment of depression}\label{sec:ex_bin}

Binomial or binary endpoints, respectively, are most commonly used in non-inferiority trials (Lange \& Freitag, 2005). In this section we introduce a clinical trial in the treatment of depression from Goldstein et al. (2004), which was also used by Kieser \& Friede (2007) for illustration. We will find in particular different answers concerning the planning of this study (see Section \ref{ex:bin2}).This randomized, double-blind trial compares duloxetine (\textbf{T}est treatment) to paroxetine (\textbf{R}eference treatment) and \textbf{P}lacebo with regard to efficacy and safety. In the therapy of depression, achieving remission is the clinically desired goal (Nierenberg \& Wright, 1999), whereas remission is defined as maintaining the Hamilton rating scale score of depression (HAM-D) total score at $\leq 7$. Table \ref{ex1:table} displays for each group, $k=T,R,P$, the total numbers of patients and the fractions of patients, who achieved remission at week 8 (end of treatment).  

{\footnotesize

\begin{table}[h]
\caption{Three-armed clinical trial in treatment of depression}\label{ex1:table}
\begin{tabular}{|l|@{\hspace{0.5cm}}r@{\hspace{0.5cm}}r@{\hspace{0.5cm}}r|}
\hline & & & \\[-0.2cm]
& & No. of Patients& Fraction of patients \\
\textbf{Treatment} & No. of patients & achieved remission & achieved remission\\
\hline\hline & & &\\[-0.2cm]
\textbf{P}lacebo & 88 & 26 & 29.55\% \\[1mm]
\textbf{R}eference& 84 & 31& 36.90\%  \\[1mm]
\textbf{T}est& 86 & 43& 50.00\% \\[1mm]
\hline
\end{tabular}
\end{table}  }

For demonstrating that duloxetine is non-inferior to the reference treatment paroxetine, following Kieser \& Friede (2007), we consider the retention of effect hypothesis with $h(\pi_k)=\pi_k$
\begin{eqnarray}\label{eqn:hyp_bin}
H_{0,\pi_k}:\;\;\pi_T-\pi_P\;\leq \;\Delta \cdot (\pi_R-\pi_P)\:,
\end{eqnarray} 
where $\pi_k$ represents the remission probability of treatment $k=T,R,P$ at the end of treatment. 

\subsection{Poisson endpoints: Treatment of epilepsy}\label{sec:ex_pois}
Typical examples of Poisson distributed endpoints can be found for example in the treatment of angina pectoris, nausea and epilepsy, see Layard \& Arvesen (1978), where the number of attacks are counted within a specified time interval, or in the treatment of depressions, where the (waiting) time until healing or remission is observed (see e.g. Mielke et al., 2008). Here, we reconsider the randomized, double blind cross-over trial in the treatment of epilepsy from Sander et al. (1990), which compares a new treatment (lamotrigine) as an add-on treatment to a placebo add-on by means of 18 patients. Table \ref{ex:table} presents the total number of seizures within the treatment weeks 9-12. Note, that Mohanraj \& Brodie (2003) highlight that for evaluating anti-epileptic drugs (AED) as add-on treatment the standard endpoint is the manipulation in the number of seizures. 

{\footnotesize
\begin{table}[h]
\caption{Three-armed clinical trial in treatment of epilepsy}\label{ex:table}
\begin{tabular}{|l|@{\hspace{0.5cm}}r@{\hspace{0.5cm}}r@{\hspace{0.5cm}}r|}
\hline & & & \\[-0.2cm]
& & & Mean no. of seizures \\
\textbf{Treatment} & No. of Patients & Total no. of seizures & per patient\\
\hline\hline & & & \\[-0.2cm]
\textbf{P}lacebo add-on& 18 & 338 & 18.78 \\[1mm]
\textbf{R}eference add-on& 18 & 295& 16.39  \\[1mm]
\textbf{T}est add-on& 18 & 288& 16.00 \\[1mm]
\hline
\end{tabular}
\end{table}  
}

As AED trials performed in the past are two-armed, either placebo- or active-controlled (for an overview see Mohanraj \& Brodie, 2003), we add for illustration purposes of our procedures an artificial reference treatment group with equal size of 18 patients and seizures of same order of magnitude as seizures under the test treatment, also displayed in Table \ref{ex:table}.

We presume that the number of seizures of each patient follows a Poisson distribution determined by the group affiliation (T,R,P), i.e. the observations are from $X_{k1},\ldots,X_{kn_k} \stackrel{i.i.d.}{\sim} Pois(\lambda_k)$ for $k=T,R,P$ with $n_P=n_T=n_R=18$. Table \ref{ex:table} displays the total number of seizures in each group, $X_k=\sum_{i=1}^{n_k} X_{ki}$, $k=T,R,P$. As in this setting small values of $\lambda_k$, representing less seizures, are desired we choose $h(\lambda_k)=-\lambda_k$, which yields the retention of effect hypothesis

\begin{eqnarray}\label{eqn:hyp_pois}
H_{0,-\lambda_k}:\; \lambda_P - \lambda_T \;\leq\; \Delta \cdot (\lambda_P-\lambda_R)\;
\end{eqnarray} 
for demonstrating that the test treatment is non-inferior to the reference one.

\subsection{Further examples}

In Table \ref{table1} we summarize various endpoints together with some common retention of effect hypotheses. Moreover, we have included some models which have not been used in the context of retention of effect hypothesis, including the Weibull- and Gamma-family. However, these endpoints are of practical interest as recent non-inferiority trials by Yakhno et al. (2006) and Gurm et al. (2008) highlight. We will not discuss all these situations in detail, but we mention that our methodology immediately applies to these situations. 

{\footnotesize
\begin{table}[h]
\caption{Survey of retention of effect hypotheses}\label{table1}
\begin{tabular}{|l|@{\hspace{0.5cm}}r@{\hspace{1.5cm}}r@{\hspace{1.5cm}}r|}
\hline & & &\\ 
\textbf{Distribution} & $\theta_k$ & $h(\theta_k)$ & $\sigma^2_k$\\
\hline\hline & & &\\[-0.2cm]
Normal (Pigeot et al., 2003) & $(\mu_k,\tau^2)$ & $\mu_k$ & $\tau^2$\\[1mm]
Normal (Hasler et al., 2008) & $(\mu_k,\tau_k^2)$ & $\mu_k$ & $\tau_k^2$\\[1mm]
Binary  & $\pi_k$ & $\pi_k$ & $\pi_k (1-\pi_k)$\\
(Kieser \& Friede, 2007, this work) & & & \\[1mm]
Binary & $\pi_k$ & $\log(\pi_k/(1-\pi_k))$ & $(\pi_k (1-\pi_k))^{-1}$\\[1mm]
Exponential (Mielke et al., 2008)& $\lambda_k$ & $\log\lambda_k$ & 1\\[1mm]
Poisson (this work)& $\lambda_k$ & $-\lambda_k$ & $\lambda_k$\\[1mm]
Gamma & $(\alpha,\beta_k)$ & $\alpha\cdot \beta_k \;[={\rm E}X_k]$& $\beta_k^2\alpha^{-1}$\\[1mm]
Weibull & $(\lambda_k,\beta)$ & $\lambda_k \; [= {\rm E}X_k \cdot (\Gamma(1+\beta^{-1}))^{-1}] $ & $I_{11}(\lambda_k,\beta)$ \\[1mm]
\hline
\end{tabular}
\end{table}  
}

\section{Wald-type test: Theory}\label{sec:test}

In this section, we derive a Wald-type test procedure for the generalized retention of effect hypothesis (\ref{eqn:g_hyp}) introduced in Section \ref{sec:intro} and discuss the estimation of the variance with restriction to the null hypothesis. This generalizes and unifies specific results for binary, normally and exponentially distributed endpoints. Based on this, we provide the theory for sample size planning in the next Section \ref{sec:planning}.
\vspace{0.4cm}

\noindent \textbf{Model assumptions:} Let $X_{ki}$ for $i=1,\ldots,n_k$ be independently distributed according to a parametric family of distributions with densities $\left\{f(\te,\cdot):\te\in\Te\right\}$, $\Te \subseteq \mathbb{R}^d$, and parameters $\te_k\in\Theta$, $k=T,R,P$, where $T,R$ and $P$ abbreviates test, reference and placebo group, respectively. We presume that the family of probability densities $\left\{f(\te,\cdot):\te\in\Te\right\}$ is sufficiently regular to obtain asymptotic normality of the ML-estimators (MLE) of the parameter $\theta$ with non-singular covariance or Fisher-information matrix, respectively, e.g. an exponential family or a family which is differentiable in quadratic mean (van der Vaart, 1998).
Moreover, none of the groups should vanish asymptotically, i.e. for $k=T,R,P$ and $n=n_T+n_R+n_P$
\begin{eqnarray}\label{eqn:propor}
\frac{n_k}{n} \longrightarrow w_k
\end{eqnarray}
holds for $n_R,n_T,n_P \rightarrow \infty$ and some $w_k\in\;]0,1[$, the (asymptotic) proportion of the numbers of patients in group $k=T,R,P$.

\subsection{Retention of Effect Wald-type Test (RET)}\label{RET_proc} 
In order to come up with a test for (\ref{eqn:g_hyp}) we rewrite this as
\begin{eqnarray}\label{eqn:hyp}
H_{0,h(\theta_k)}:\;\; \eta:=\; h(\te_T)-\Delta\:h(\te_R)+(\Delta-1)h(\te_P)\;\;\leq \;\; 0\: . 
\end{eqnarray}
The MLE of $h(\te_k)$, $k=T,R,P$, is obtained by plugging in the MLE $\hat{\te}_k$ of $\te_k$, which is well-defined and asymptotically normally distributed by assumption. By the delta-method this yields that $\sqrt{n_k}(h(\hat{\te}_k)-h(\te_k))$ is centered asymptotically normally distributed with variance 
$$
\sigma^2_k=\left(\frac{\partial}{\partial \te}h(\te_k)\right)\cdot I(\theta_k)^{-1}\cdot\left(\frac{\partial}{\partial \te}h(\te_k)\right)^T
$$
and $I$ the Fisher-information-matrix, i.e.
$$
I(\theta) = -E_{\te}\left [ \frac{\partial^2}{\partial^2 \te} \log f(\te,X) \right ]\:.
$$
Hence, the linear contrast $\sqrt{n}(\hat{\eta}-\eta)$, where the MLE of $\eta$ is obtained by plugging in the MLE's $\hat{\te}_k$, $k=R,T,P$, in the left hand side of (\ref{eqn:hyp}), is centered asymptotically normal with variance
\begin{eqnarray}\label{eqn:var}
\sigma^2 \; = \; \frac{\sigma^2_T}{w_T} + \frac{\Delta^2 \sigma^2_R}{w_R} + \frac{(1-\Delta)^2\sigma^2_P}{w_P}\;. 
\end{eqnarray}
As we have mentioned in the introduction estimation of $\sigma^2$ simply by the MLE often leads to an unsatisfactory approximation of the asymptotic normal law and various improvements have been suggested in specific settings, mainly for the case of binary endpoints (see next section). Therefore, we will treat the case of restricted maximum likelihood estimation as well. To this end let $\hat{\sigma}^2_{ML}$ denote the MLE of $\sigma^2$ and $\hat{\sigma}^2_{RML}$ denote the MLE with restriction to the null hypothesis, i.e. the MLE of $\sigma^2$ under the restriction in (\ref{eqn:hyp}). Further let $\hat{\sigma}^2$ either denote $\hat{\sigma}^2_{ML}$ or $\hat{\sigma}^2_{RML}$, see the next Section \ref{comp_sigma} for a discussion of both estimators. Both estimators are consistent under the null hypothesis. Thus, we obtain in order to test (\ref{eqn:hyp}) as a test-statistic
\begin{eqnarray}\label{eqn:teststatistic}
T=\frac{\sqrt{n}\cdot\hat{\eta}}{\hat{\sigma}}=\sqrt{n}\cdot\frac{h(\hat{\te}_T)-\Delta\:h(\hat{\te}_R)+(\Delta-1)h(\hat{\te}_P)}{\hat{\sigma}}
\end{eqnarray}
which is asymptotically standard normally distributed at the boundary of $H_{0,h(\theta_k)}$, i.e. when $\eta=0$. Therefore, $H_{0,h(\theta_k)}$ is to reject if $T>z_{1-\alpha}$, where $z_{1-\alpha}$ is the $1-\alpha$-quantile of the standard normal distribution and $\alpha$ a specified significance level. Due to the formulation of the hypothesis and the test decision we will denote this test by \textit{Retention of Effect Wald Test (RET)}. 

\subsection{The estimators of the asymptotic variance $\sigma^2$ and their limits}
In some situation, e.g. for normally distributed endpoints, it is sufficient to estimate the asymptotic variance in (\ref{eqn:var}) by the (unrestricted) MLE (see Pigeot et al., 2003). Roughly speaking, this is due to the fact that the asymptotic variance of the test statistic does not depend on the parameters $h(\theta_k)$ (in the normal case the mean) which only enter into the hypothesis. However, e.g. for the case of binary endpoints the variance depends on the success probabilities itself and an improvement in the accuracy of the asymptotic normality can be obtained by estimation restrictedly to the null hypothesis. This has been pointed out by Farrington \& Manning (1990) for the two sample comparison with binomial endpoints and various improvements have been suggested since (see e.g. Chan (1998), R\"ohmel \& Mansmann (1999), Skipka et al. (2004)). For the retention of effect hypothesis Kieser \& Friede (2007) demonstrate in an extensive simulation study that the restricted Wald-type test (Farrington \& Manning's (1990) adjustment) works satisfactorily and clearly outperforms the unrestricted Wald-type test concerning the accuracy of the nominal level. 

\subsubsection{Computation of $\hat{\sigma}^2_{ML}$ and $\hat{\sigma}^2_{RML}$}\label{comp_sigma}
Typically, the variance $\sigma^2$ is a continuous function of the parameters $\theta_k$, $k=T,R,P$. Thus, the MLE $\hat{\sigma}^2_{ML}$ is obtained by plugging the MLE's $\hat{\theta}_k$
$$
\hat{\sigma}_{ML}=\sigma(\hat{\theta}_T,\hat{\theta}_R,\hat{\theta}_P).
$$
In order to obtain the restricted MLE $\hat{\sigma}^2_{RML}$ the $\hat{\theta}_k$'s have to be replaced by their restricted versions, i.e. 
$$
\hat{\sigma}_{RML}=\sigma(\hat{\theta}_{T,H_0},\hat{\theta}_{R,H_0},\hat{\theta}_{P,H_0})
$$
with
\begin{eqnarray}\label{likelihood}
(\hat{\theta}_{T,H_0},\hat{\theta}_{R,H_0},\hat{\theta}_{P,H_0})= {\arg\sup}_{(\theta_{T},\theta_{R},\theta_{P})\in H_{0,h(\theta_k)} } \;\sum_{k=T,R,P} \sum_{i=1}^{n_k}\log f(\theta_k,x_{ki}) .
\end{eqnarray}

The restricted MLEs $(\hat{\theta}_{T,H_0},\hat{\theta}_{R,H_0},\hat{\theta}_{P,H_0})$ can be computed in the following way: if the unrestricted MLEs $\hat{\theta}_k$, $k=T,R,P$, are located in $H_{0,h(\theta_k)}$, i.e. $h(\hat{\theta}_T)-\Delta\:h(\hat{\theta}_R)+(\Delta-1)h(\hat{\theta}_P)\leq0$, they coincide with the restricted MLEs. Otherwise the restricted MLEs can be determined by restricting the likelihood function to the boundary of $H_{0,h(\theta_k)}$ by means of substituting $\theta_T=h^{-1}(\Delta h(\theta_R)+(1-\Delta) h(\theta_P))$ in the common likelihood function (left hand side from (\ref{likelihood})) and maximizing this with respect to $\theta_R$ and $\theta_P$ numerically or, if possible, analytically.

\subsubsection{Limits of the variance estimators} 
The limits of the MLEs $\hat{\sigma}^2_{ML}$ and $\hat{\sigma}^2_{RML}$ are crucial for sample size planning in the following Section \ref{sec:planning}. For the derivation of the limits let us denote the true (unknown) parameters by $\theta_k^{(0)}$, $k=T,R,P$,  and correspondingly $\eta^{(0)}= h(\te_T^{(0)})-\Delta\:h(\te_R^{(0)})+(\Delta-1)h(\te_P^{(0)})$ and
\begin{eqnarray}\label{sigma_0k}
\sigma_{0,k}^2=\left(\frac{\partial}{\partial \te}h(\te_k^{(0)})\right)\cdot I(\theta_k^{(0)})^{-1}\cdot\left(\frac{\partial}{\partial \te}h(\te_k^{(0)})\right)^T
\end{eqnarray}
for $k=R,T,P$ and 
\begin{eqnarray}\label{var_0}
\sigma^2_0 \; = \; \frac{\sigma^2_{0,T}}{w_T} + \frac{\Delta^2 \sigma^2_{0,R}}{w_R} + \frac{(1-\Delta)^2\sigma^2_{0,P}}{w_P}\;. 
\end{eqnarray}
The unrestricted MLE $\hat{\sigma}^2_{ML}$ is always a consistent estimator, i.e. $\hat{\sigma}^2_{ML}\stackrel{a.s.}{\longrightarrow}\sigma^2_0$ as $n\rightarrow\infty$. However, the restricted MLE $\hat{\sigma}^2_{RML}$ is only consistent when the true parameters are located in the hypothesis, i.e.  $\eta^{(0)}\leq0$. In other words, the limit of $\hat{\sigma}^2_{RML}$ is no more equal to $\sigma^2_0$, in general. We will now derive the limit of the restricted MLE $\hat{\sigma}^2_{RML}$, when the parameters are located in the alternative, i.e. $\eta^{(0)}>0$. This requires computation of the Kullback-Leibler-divergence (KL-divergence) between two parameter constellations. To this end, let $\zeta=(\theta_T,\theta_R,\theta_P)$ denote any parameter in the parameter space $\Theta^3\subseteq \mathbb{R}^{3d}$ and $\zeta^{(0)}$ the true parameter. Then we define for the three-sample case a weighted KL-divergence between $\zeta^{(0)}$ and $\zeta$ with weights $c=(c_T,c_R,c_P)$ by
\begin{eqnarray}\label{w_KL}
K(\zeta^{(0)},\zeta,c)= \sum_{k=T,R,P} c_k \cdot K(\theta_k^{(0)},\theta_k) \;,
\end{eqnarray}
where $K(\theta_k^{(0)},\theta_k)=E_{\theta_k^{(0)}}[\log f(\theta_k^{(0)},X)-\log f(\theta_k,X)]$ denotes the usual KL-divergence measuring the difference between two densities. According to Theorem 2 (see Appendix \ref{appendix:proof}) the restricted MLE $\hat{\zeta}_{H_0}=(\hat{\theta}_{T,H_0},\hat{\theta}_{R,H_0},\hat{\theta}_{P,H_0})$ converges to the minimizer of the sample size weighted KL-divergence to the true parameter, i.e. 
$$
\hat{\zeta}_{H_0} \stackrel{a.s.}{\longrightarrow} \zeta_{H_0}
$$
with
$$
\zeta_{H_0}=(\theta_{T,H_0},\theta_{R,H_0},\theta_{P,H_0})=\arg\min_{\zeta\in H_0} K(\zeta^{(0)},\zeta,(w_T,w_R,w_P)).
$$
Therefore, the limit of the restricted MLE $\hat{\sigma}^2_{RML}$ is obtained by
\begin{eqnarray}\label{s_rml}
\sigma_{RML}^2 \; = \; \frac{\sigma^2_{T,H_0}}{w_T} + \frac{\Delta^2 \sigma^2_{R,H_0}}{w_R} + \frac{(1-\Delta)^2\sigma^2_{P,H_0}}{w_P}
\end{eqnarray}
with
\begin{eqnarray}\label{s_H0}
\sigma^2_{k,H_0}=\left(\frac{\partial}{\partial \te}h(\te_{k,H_0})\right)\cdot I(\theta_{k,H_0})^{-1}\cdot\left(\frac{\partial}{\partial \te}h(\te_{k,H_0})\right)^T
\end{eqnarray}
for $k=T,R,P$.

\subsubsection{Numerical computation of $\sigma_{RML}$}\label{sec:KL-practice} 
For computing the minimizers $\theta_{k,H_0}$, $k=T,R,P$, and therewith $\sigma_{RML}$ for a parameter constellation in the alternative, i.e. $\eta^{(0)}>0$, it is sufficient to restrict to the boundary of $H_{0,h(\theta_k)}$, i.e. we replace in the weighted KL-divergence (\ref{w_KL}) $\theta_{T}$ by $h^{-1}(\Delta h(\theta_R)+(1-\Delta)h(\theta_P))$ and then minimize the KL-divergence with respect to $\theta_R$ and $\theta_P$. 

In practice, the analytic solution to the minimization problem of the KL-divergence may be hard (confer the example of Poisson endpoints in Section \ref{plan_pois}) or even unfeasible to find. In this case, numerical minimization becomes necessary. To this end, it is important to note that the minimization of the KL-divergence often results in a convex optimization problem and fast algorithms for convex optimization , such as the Newton-Raphson algorithm, become feasible. The following theorem states conditions to obtain a convex optimization problem.

\smallskip

\noindent \textbf{Theorem 1:} \textit{Let $-E_{\theta_k^{(0)}}[\frac{\partial^2}{\partial^2\theta}\log f(\theta,X)]$ be non-negative for all $\theta\in\Theta$ and $\theta_{k}^{(0)}$, $k=T,R,P$ and let $\Theta$ be a convex set. Further, let $h^{-1}(\Delta h(\theta_R)+(1-\Delta)h(\theta_P))$ be an affine transformation in $\theta_R$ and $\theta_P$. Then, restricted to the boundary of the null hypothesis, the minimization in $\zeta$ of the weighted KL-divergence (\ref{w_KL}) is a convex optimization problem.}

\smallskip

The conditions of Theorem 1 are fulfilled in our examples of Poisson and binary endpoints, which will be revisited in Section \ref{sec:examples_revisited}.

\subsection{Approximating the power function of the RET}

The asymptotic normality used in Section \ref{RET_proc} to derive the RET is valid for parameter constellations in the hypothesis as well as for constellations in the alternative. Thus, if the variance $\sigma^2$ is estimated unrestrictedly, $\hat{\sigma}^2=\hat{\sigma}^2_{ML}$, we obtain as an approximation to the power function of the RET, i.e. the probability of rejecting the hypothesis $H_{0,h(\theta_k)}$ in (\ref{eqn:g_hyp}), by
\begin{eqnarray}\label{eqn:power_unres}
P_{\eta^{(0)}}\left(T > z_{1-\alpha} \right) \approx  1- \Phi\left(z_{1-\alpha}-\sqrt{n}\;\frac{\eta^{(0)}}{\sigma_0} \right)\;.
\end{eqnarray}
However, estimating the variance $\sigma^2$ restricted to the null hypothesis, i.e. $\hat{\sigma}^2=\hat{\sigma}^2_{RML}$, complicates the issue and changes the power function to
\begin{eqnarray}\label{eqn:power}
P_{\eta^{(0)}}\left(T > z_{1-\alpha} \right)&=&P_{\eta^{(0)}}\left(T\cdot \frac{\sigma_{RML}}{\sigma_0}-\sqrt{n}\;\frac{\eta^{(0)}}{\sigma_0} > z_{1-\alpha} \cdot \frac{\sigma_{RML}}{\sigma_0}-\sqrt{n}\;\frac{\eta^{(0)}}{\sigma_0} \right) \nonumber \\ 
&\approx & 1- \Phi\left(z_{1-\alpha} \cdot \frac{\sigma_{RML}}{\sigma_0}-\sqrt{n}\;\frac{\eta^{(0)}}{\sigma_0} \right)\:.
\end{eqnarray}
Note, that (\ref{eqn:power_unres}) can be obtained from (\ref{eqn:power}) by means of substituting $\sigma_{RML}$ by $\sigma_{0}$.
 
\section{Sample size formula and optimal allocation of samples}\label{sec:planning}
 
In this section, we present a sample size formula for the test of the generalized retention of effect hypothesis $H_{0,h(\theta_k)}$ (\ref{eqn:g_hyp}) introduced in Section \ref{sec:intro}. In particular, we derive the optimal allocation of the samples to the groups $T,R$ and $P$ in terms of maximizing the power of the RET under any fixed alternative $\eta^{(0)}$.
 
\subsection{Optimal sample allocation}

In planning a trial, one typically specifies a parameter constellation $\eta^{(0)}$ in the alternative. Our aim in this section is to optimize the allocation of samples, represented through $w_k$, $k=T,R,P$, as in (\ref{eqn:propor}), such that the power of the test decision in (\ref{eqn:power_unres}) or (\ref{eqn:power}), respectively, is maximized. The power depends on the allocation through $\sigma^2_0$ and $\sigma^2_{RML}$.

When the variance $\sigma^2$ is estimated unrestricted in the test procedure ($\hat{\sigma}^2=\hat{\sigma}^2_{ML}$) we only have to consider $\sigma^2_0$ to investigate the influence of the allocation on the power, confer (\ref{eqn:power_unres}). This means that we have to minimize $\sigma^2_0$ in order to maximize the power. By straight forward calculations, presented in Appendix \ref{app:min_var}, we obtain as major and general valid result that the (asymptotically) optimal allocation of samples for the RET is given by
\begin{eqnarray}\label{opt_alloc}
n_T^*:n_R^*:n_P^*\;=\;1:\Delta\:\frac{\sigma_{0,R}}{\sigma_{0,T}}:|1-\Delta|\:\frac{\sigma_{0,P}}{\sigma_{0,T}}.
\end{eqnarray}
The resulting optimal minimal variance is given by
$$
\sigma^2_{0,optimal}=\left(\sigma_{0,T}+\Delta\sigma_{0,R}+|1-\Delta|\sigma_{0,P} \right)^2\:.
$$
\noindent\textbf{Remark:} For the specific case of normal endpoints with equal variances (Pigeot et al. (2003), Schwartz \& Denne (2006)) and exponentially distributed endpoints (Mielke et al., 2008) we obtain the optimal allocation $1:\Delta:|1-\Delta|$, again.

\smallskip
When the variance $\sigma^2$ is estimated under restriction to $H_{0,h(\theta_k)}$ the asymptotic power in (\ref{eqn:power}) depends additionally on $\sigma_{RML}/\sigma_0$ because under any alternative the restricted estimator $\hat{\sigma}_{RML}$ is not a consistent estimator for $\sigma_0$. Nevertheless, the asymptotically optimal allocation derived for the unrestricted case is again optimal in an asymptotic sense because the power in (\ref{eqn:power}) is dominated by the term $\sqrt{n}\cdot\eta{(0)}/\sigma_0$ as $n$ grows. Hence, the allocation (\ref{opt_alloc}) derived in the previous section, which minimizes the variance $\sigma_0$, is also the (asymptotically) optimal allocation in terms of maximizing the power when the variance $\sigma^2$ is estimated restricted to $H_{0,h(\theta_k)}$.

\smallskip

\noindent\textbf{Remark:} (a) We would like to stress that this result can be applied to the case of binary endpoints (see Section \ref{sec:bin_opt}). This leads to different results as in Kieser \& Friede (2007), who derived the optimal allocation under the additional restriction of a fixed ratio $w_R/w_T$. 

\noindent (b) The asymptotically optimal allocation presented in (\ref{opt_alloc}) should be understood as approximative for finite samples as it is customary for asymptotic results. Nevertheless, for the presented examples in this paper we will show in Section \ref{sec:examples_revisited} that the optimal allocation is also very accurate for finite samples, e.g. for a power of 80\%. However, one should be aware of the fact that, in particular for small sample sizes, it is not guaranteed that the allocation (\ref{opt_alloc}) is optimal, in general. 

\subsubsection{Rule of thumb}\label{RuleOfThumb}

The asymptotically optimal sample allocation (\ref{opt_alloc}) depends on the choice of the alternative $\zeta^{(0)}>0$. If one is not clear about the choice of the alternative or wants to consider more than one alternative, we recommend to use as a rule of thumb the allocation $1:\Delta:(1-\Delta)$. We will show for $\theta_R^{(0)}=\theta_T^{(0)}$ and $0\leq\Delta\leq 1$ in the Appendix \ref{sec:comp_var} that the allocation $1:\Delta:(1-\Delta)$ is more appropriate than the commonly used 2:2:1 allocation (the balanced allocation) if $\sigma^2_{0,P}/\sigma^2_{0,T}<2.12 (2.73)$. Note, that a lower bound for $\sigma^2_{0,P}/\sigma^2_{0,T}$ is not required. Moreover, this result is valid independent of the distribution of the endpoints and of the formulation of the retention of effect hypothesis.

\subsection{Sample size computation}\label{ss_comp}

When the variance $\sigma^2$ is estimated unrestrictedly ($\hat{\sigma}^2=\hat{\sigma}^2_{ML}$) we end up with the simplified power formula (\ref{eqn:power_unres}). Thus,  the minimal required total sample size to obtain a power of $1-\beta$ for a given alternative $\eta^{(0)}>0$ is determined by
\begin{eqnarray}\label{samp_for_1}
n_{1-\beta}\;\approx\; \left(z_{1-\alpha}+z_{1-\beta} \right)^2\cdot \left(\frac{\sigma_0}{\eta^{(0)}}\right)^2  
\end{eqnarray} 
with $\sigma_0$ defined in (\ref{var_0}). When the variance $\sigma^2$ is estimated restricted to the null hypothesis ($\hat{\sigma}^2=\hat{\sigma}^2_{RML}$) the sample size formula has to be derived from (\ref{eqn:power}) and becomes more involved, viz.
\begin{eqnarray}\label{eqn:samplesize}
n_{1-\beta}\;\approx\;\left(\frac{z_{1-\alpha}\cdot\sigma_{RML}+z_{1-\beta}\cdot\sigma_0}{\eta^{(0)}}  \right) ^2 =\left(z_{1-\alpha}\cdot\frac{\sigma_{RML}}{\sigma_0}+z_{1-\beta} \right)^2\cdot \left(\frac{\sigma_0}{\eta^{(0)}}\right)^2  ,
\end{eqnarray}
with $\sigma_{RML}$ derived in (\ref{s_rml}). As we will see the additional term $\sigma_{RML}/\sigma_0$ has a relevant impact on the sample size planning. 

In Figure \ref{algo_samplssize} we have summarized the general strategy for sample size planning (GSSP) for the RET when the variance $\sigma^2$ is estimated with restriction to the null hypothesis. When the variance $\sigma^2$ is estimated unrestrictedly by $\hat{\theta}_{ML}$ we may omit the steps 2.-4. in Figure \ref{algo_samplssize} and use the simpler formula (\ref{samp_for_1}) in step 5. to compute the required sample size $n_{1-\beta}$.

{\footnotesize
\begin{figure}[t]
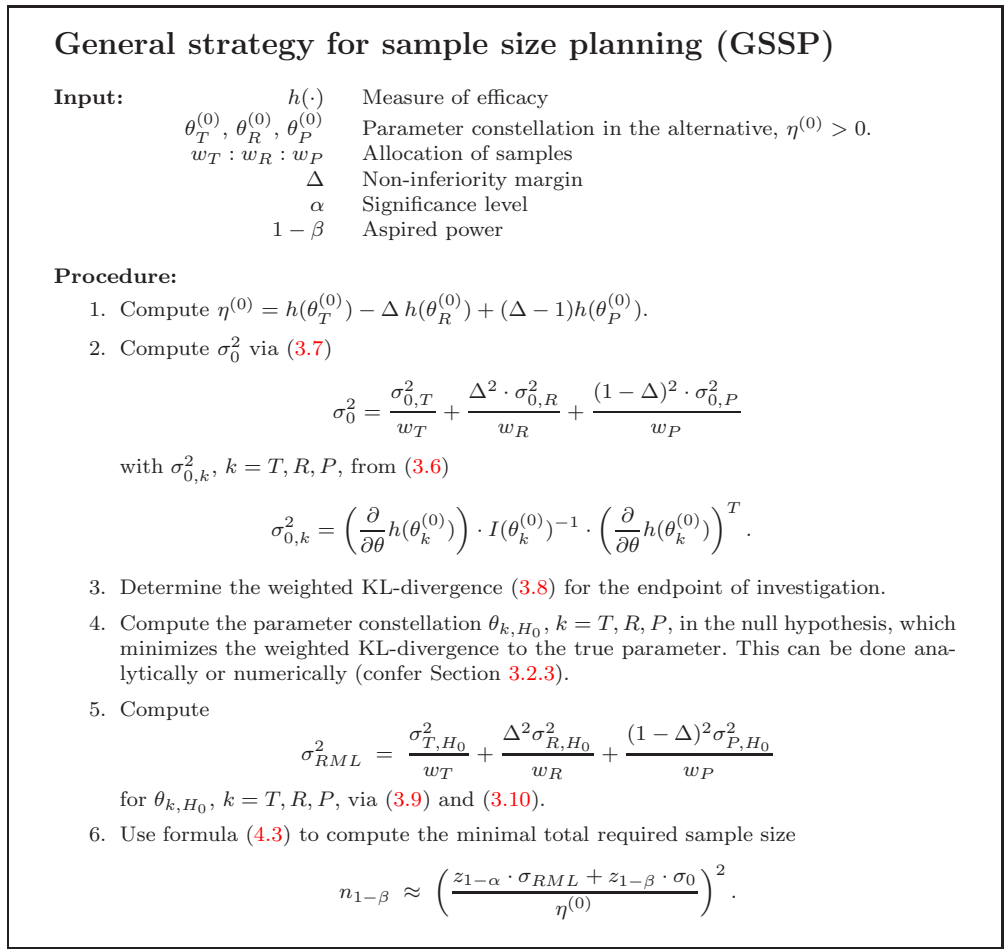

\centering
\fbox{\hspace{0.5cm}\parbox{12cm}{
\vspace{0.2cm}
\textbf{\large General strategy for sample size planning (GSSP)}\\[0.3cm]
\textbf{Input:} \hspace{0.5cm}\begin{tabular}[t]{r@{\hspace{0.5cm}}l}
$h(\cdot)$& Measure of efficacy \\
$\theta_T^{(0)}$, $\theta_R^{(0)}$, $\theta_P^{(0)}$ & Parameter constellation in the alternative, $\eta^{(0)}>0$.\\
$w_T:w_R:w_P$ & Allocation of samples\\
$\Delta$ & Non-inferiority margin\\
$\alpha$ & Significance level\\
$1-\beta$ & Aspired power\\
\end{tabular}
\vspace{0.3cm}

\textbf{Procedure:}
\begin{enumerate} 
\item Compute $\eta^{(0)}= h(\te_T^{(0)})-\Delta\:h(\te_R^{(0)})+(\Delta-1)h(\te_P^{(0)})$.
\item Compute $\sigma_0^2$ via (\ref{var_0})
$$
\sigma_0^2 = \frac{\sigma_{0,T}^2}{w_T}+\frac{\Delta^2 \cdot\sigma_{0,R}^2}{w_R}+\frac{(1-\Delta)^2\cdot\sigma_{0,P}^2}{w_P}
$$
with $\sigma_{0,k}^2$, $k=T,R,P$, from (\ref{sigma_0k})
\begin{eqnarray*}
\sigma_{0,k}^2=\left(\frac{\partial}{\partial \te}h(\te_k^{(0)})\right)\cdot I(\theta_k^{(0)})^{-1}\cdot\left(\frac{\partial}{\partial \te}h(\te_k^{(0)})\right)^T.
\end{eqnarray*}
\item Determine the weighted KL-divergence (\ref{w_KL}) for the endpoint of investigation. 
\item Compute the parameter constellation $\theta_{k,H_0}$, $k=T,R,P$, in the null hypothesis, which minimizes the weighted KL-divergence to the true parameter. This can be done analytically or numerically (confer Section \ref{sec:KL-practice}). 
\item Compute 
$$
\sigma_{RML}^2 \; = \; \frac{\sigma^2_{T,H_0}}{w_T} + \frac{\Delta^2 \sigma^2_{R,H_0}}{w_R} + \frac{(1-\Delta)^2\sigma^2_{P,H_0}}{w_P}
$$  
for $\theta_{k,H_0}$, $k=T,R,P$, via (\ref{s_rml}) and (\ref{s_H0}).
\item Use formula (\ref{eqn:samplesize}) to compute the minimal total required sample size
$$
n_{1-\beta}\;\approx\;\left(\frac{z_{1-\alpha}\cdot\sigma_{RML}+z_{1-\beta}\cdot\sigma_0}{\eta^{(0)}}  \right) ^2 .
$$
\end{enumerate}
}\hspace{0.5cm}}
\caption{General strategy for sample size planning (GSSP) when the variance $\sigma^2$ is estimated with restriction to the null hypothesis.} \label{algo_samplssize}
\end{figure}   
}

\smallskip

\noindent \textbf{Remark:} We stress again that the use of $\sigma_{RML}$ will affect the planning of the trial significantly. If one replaces in (\ref{eqn:samplesize}) $\sigma_{RML}$ by $\sigma_{0}$ this may result in a too small or too large required sample size depending on the ratio $\sigma_{RML}/\sigma_{0}$. If the ratio $\sigma_{RML}/\sigma_{0}$ is greater (smaller) than one, then we end up with a too small (large) required sample size, i.e. the resulting power is smaller (larger) than the desired power $1-\beta$. For example, this will be the case for Poisson distributed endpoints (see Section \ref{sec:ex_pois}) and the hypothesis (\ref{eqn:hyp_pois}). We will see in Section \ref{ex:pois2} that $\sigma_{RML}/\sigma_{0}>1$ for all parameter constellations. In contrast, for binary endpoints (see Section \ref{sec:ex_bin}) and the hypothesis (\ref{eqn:hyp_bin}), we will show in Section \ref{ex:bin2} that there is no strict relationship between $\sigma_{RML}$ and $\sigma_{0}$. Thus, a wrongly specified sample size may result in a too large or too small power compared to the aspired one.

\section{Examples revisited}\label{sec:examples_revisited}

In the following  we will perform the RET for the examples from Section \ref{sec:example} and we will illustrate the general strategy for sample size planning (GSSP) including a detailed investigation of the optimal allocation. 

\subsection{Binary endpoints: Treatment of depression}\label{ex:bin2}

In this section, we revisit the example in the treatment of depression introduced in Section \ref{sec:ex_bin}. 

\begin{figure}[t]
\centering
\includegraphics[height=6cm, angle=-90]{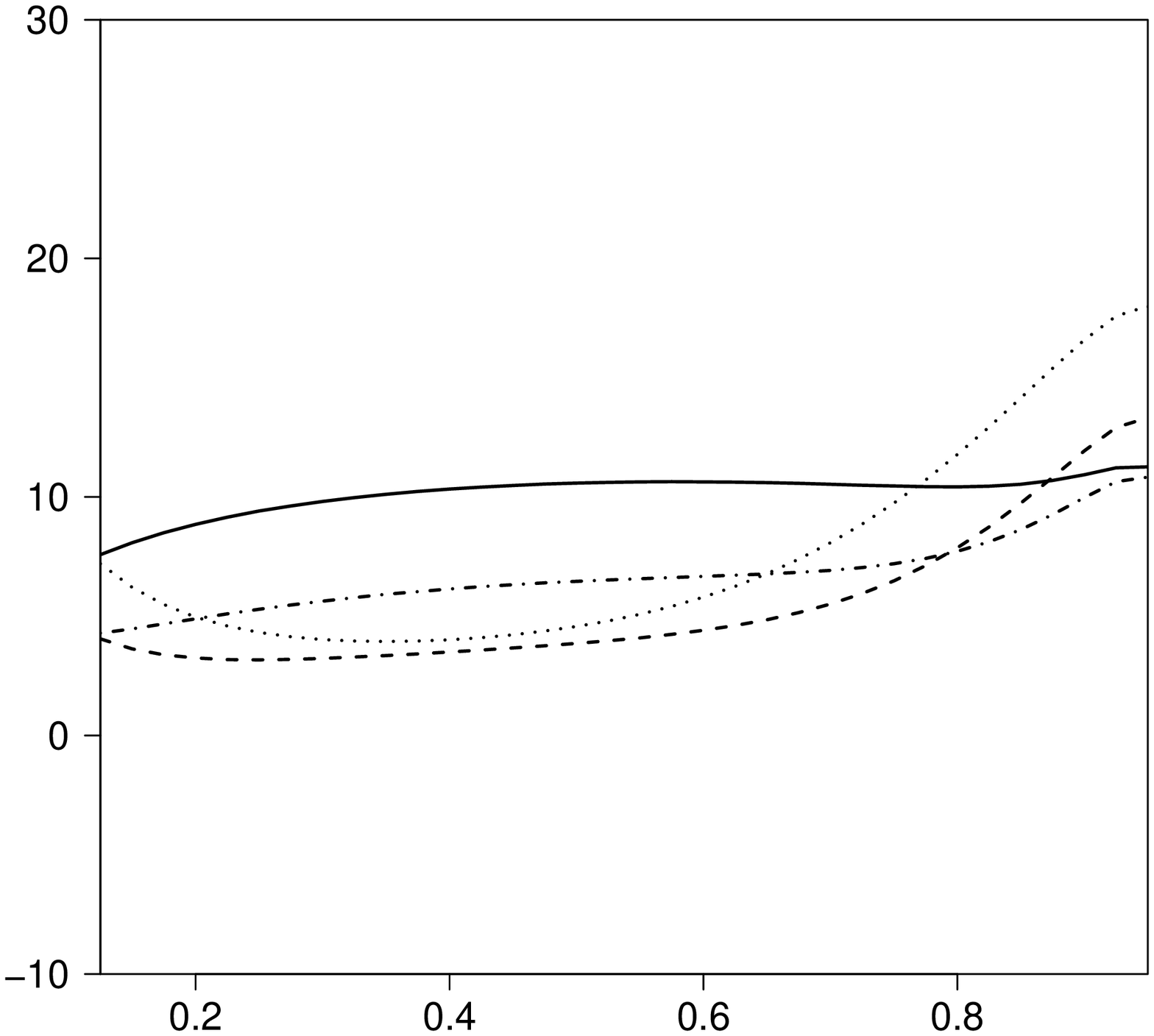}
\put(-180,-80){\%}
\put(-90,-170){$\pi_{0,R}$}
\hspace{1cm}
\includegraphics[height=6cm, angle=-90]{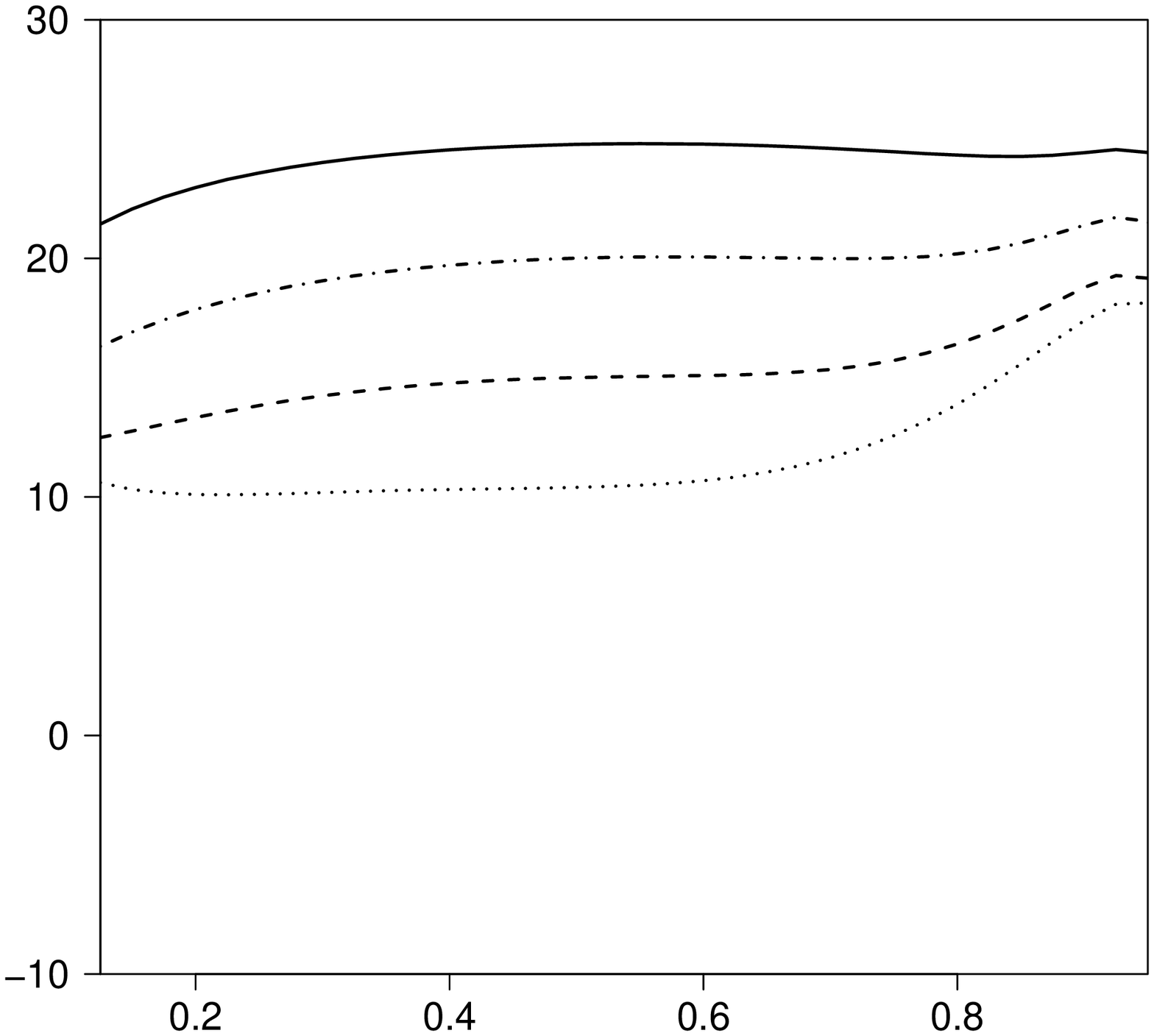}
\put(-90,-170){$\pi_{0,R}$}
\put(-180,-80){\%}

\caption{Example of binary distributed endpoints: \textbf{Sample size reduction in \%} when optimal allocation (\ref{bin_opt3}) is used instead of the balanced allocation (right figure) and instead of the allocation 2:2:1 (left figure) for $\pi_{0,P}=0.1$ and different values $\Delta=0.5$ (dotted line), $\Delta=0.6$ (dashed line), $\Delta=0.7$ (dotdash line), $\Delta=0.8$ (solid line). } \label{fig:red2}
\end{figure}

\subsubsection{Performing the RET} 
For the sake of completeness we recall the RET for the situation $h(\pi_k)=\pi_k$, which was already introduced by Tang \& Tang (2004) and Kieser \& Friede (2007). The MLE of $\pi_k$ is $\hat{\pi}_k=n_k^{-1}\sum_{i=1}^{n_k}X_{ki}$ which is asymptotically normally distributed with variance $\sigma_k^2=\pi_k(1-\pi_k)$. Hence, the unrestricted MLE of the variance $\sigma^2$ is given by (cf. (\ref{eqn:var})), 
$$
\hat{\sigma}^2_{ML}=n\cdot\left(\frac{\hat{\pi}_T(1-\hat{\pi}_T)}{n_T}+ \Delta^2\:\frac{\hat{\pi}_R(1-\hat{\pi}_R)}{n_R}+ (1-\Delta)^2\:\frac{\hat{\pi}_P(1-\hat{\pi}_P)}{n_P} \right)
$$
and we end up with the test statistic (see (\ref{eqn:teststatistic}))
\begin{eqnarray}\label{eqn:teststat_bin}
T=\sqrt{n}\cdot\frac{\hat{\pi}_T-\Delta\:\hat{\pi}_R+(\Delta-1)\hat{\pi}_P}{\hat{\sigma}_{ML}}
\end{eqnarray}
in order to test $H_{0,\pi_k}$ in (\ref{eqn:hyp_bin}), which is rejected if $T>z_{1-\alpha}$. 

Let us now consider the case where $\sigma^2$ is estimated restrictedly (cf. Farrington \& Manning, 1990). The restricted version of the Wald-type test is observed by replacing the MLEs $\hat{\pi}_k$ in the denominator by the to $H_{0,\pi_k}$ restricted ones. Here, we have computed the restricted MLEs accordingly to Section \ref{comp_sigma} by means of substituting $\pi_T=\Delta\pi_R+(1-\Delta)\pi_P$ in the common likelihood function and maximizing this with respect to $\pi_R$ and $\pi_P$ numerically. Note, that in contrast to the two-sample case (Farrington \& Manning, 1990), an analytical computation of the restricted MLE's is not feasible, anymore. 

The RET for the hypothesis (\ref{eqn:hyp_bin}) with $\Delta=0.8$ yields $T=2.104$ $(2.108)$ in (\ref{eqn:teststat_bin}) using the restricted (unrestricted) estimator for the variance estimation and corresponding p-values 1.77\% (1.75\%). Thus, we would reject $H_{0,\theta_k}$ from (\ref{eqn:hyp_bin}) in both cases and claim that the test treatment duloxetine is non-inferior to paroxetine. 

{\footnotesize
\begin{table}[h]
\caption{\textbf{Example of binomial distributed endpoints:} Optimal sample allocation, limit of variance estimator $\hat{\sigma}_{RML}$ and required samples size from formula (\ref{eqn:samplesize}) and (\ref{samp_for_1}), respectively, to obtain a power of 0.7 and 0.8, respectively, when the variance $\sigma^2$ is estimated restrictedly to the null-hypothesis (unrestrictedly), where $\alpha=5\%$, $\Delta=0.7$. } \label{table_bin}
\begin{tabular}{|lc|rrr|rrr|rrr|}
\hline & & & & & & & & & &  \\[-0.2cm]
 & & & & &  \multicolumn{3}{|c|}{Optimal allocation} & \multicolumn{3}{|c|}{2:2:1 allocation}\\
$\pi_{0,P}$ & $\pi_{0,T}$ &$w_T^*$ &$w_R^*$ &$w_P^*$ &$\frac{\sigma_{RML}}{\sigma_0}$& $n_{0.7}$ & $n_{0.8}$ & $\frac{\sigma_{RML}}{\sigma_0}$& $n_{0.7}$ & $n_{0.8}$ \\[0.2cm]
\hline\hline & & & & & & & & &  \\[-0.2cm]
0.1 &0.3 &0.527 &0.369 &0.104  &0.994 &997 (988) &1308 (1297)& 1.014 & 1054 (1076) & 1388 (1414)\\
    &0.5 &0.532 &0.372 &0.096  &0.986 &296 (289) &387 (380) & 1.006 & 315 (318) & 415 (418)\\
    &0.7 &0.527 &0.369 &0.104  &0.955 &118 (110) &154 (145) & 0.965 & 127 (120) & 165 (158)\\
    &0.9 &0.500 &0.350 &0.150  &0.791 &43 (30) & 54 (39) & 0.759 & 48 (31) & 60 (41)\\[0.1cm]
0.3 &0.5 &0.506 &0.354 &0.139  &0.998 &1279 (1275) &1680 (1675) & 1.012 & 1341 (1341) & 1761 (1762)\\
    &0.7 &0.500 &0.350 &0.150  &0.986 &281 (275) &368 (361) & 0.975 & 298 (287) & 390 (377)\\
    &0.9 &0.463 &0.324 &0.212  &0.867 &76 (61) &98 (81) & 0.830 & 84 (63) & 106 (83)\\[0.1cm]
0.5 &0.7 &0.493 &0.345 &0.161  &0.997 &1134 (1129) &1489 (1483) & 0.988 & 1191 (1170) & 1561 (1537)\\
    &0.9 &0.455 &0.318 &0.227  &0.924 &161 (143) &209 (188) & 0.894 & 174 (147) & 224 (193)\\[0.1cm]
0.7 &0.8 &0.489 &0.343 &0.168  &0.998 &3505 (3495) &4603 (4591) & 0.989 & 3672 (3611) & 4814 (4744)\\
    &0.9 &0.463 &0.324 &0.212  &0.974 &571 (549) &746 (721) & 0.949 & 609 (562) & 792 (739)\\[0.1cm]
0.8 &0.9 &0.476 &0.333 &0.190  &0.992 &2101 (2076) &2756 (2727) & 0.975 & 2214 (2130) & 2895 (2798)\\

\hline

\end{tabular}
\end{table}  
}

{\footnotesize
\begin{table}[h]
\caption{Precision of sample size formula (\ref{eqn:samplesize}) and comparison to the results obtained by Kieser \& Friede (2007) for a aspired power of 80\% at significance level $\alpha=2.5\%$. } \label{table_bin2}
\begin{tabular}{|lccc|rr|rr|}
\hline & & & & & & & \\[-0.2cm]
 & & & & & & \multicolumn{2}{c|}{Usage of Eq. (\ref{eqn:samplesize}) from this}\\
 & & & &  \multicolumn{2}{|c|}{Kieser \& Friede (2007)} & \multicolumn{2}{c|}{work with exact limit $\sigma_{RML}$}\\[0.2cm]
 $w_T:w_R:w_P$ & $\Delta$ & $\pi_{0,P}$ & $\pi_{0,R}$ & $n$ & Exact Power & $n$ & Exact Power\\[0.2cm]
 \hline\hline & & & & & & & \\[-0.2cm]
 1:1:1 & 0.6 & 0.1 & 0.5 & 309 & 78.94\% & 319 & 80.08\% \\
  		 &		 & 0.1 & 0.7 & 135 & 81.51\% & 132 & 80.77\% \\
  		 &		 & 0.1 & 0.9 & 54  & 83.05\% & 53  & 80.49\% \\
  		 &		 & 0.3 & 0.7 & 318 & 81.17\% & 312 & 80.45\% \\
  		 &		 & 0.3 & 0.9 & 99  & 83.92\% & 94  & 81.52\% \\
  		 &		 & 0.5 & 0.9 & 213 & 84.95\% & 195 & 81.43\% \\[0.1cm]
  		 & 0.8 & 0.1 & 0.7 & 606 & 81.74\% & 583 & 80.18\% \\
  		 &     & 0.1 & 0.9 & 201 & 85.57\% & 182 & 81.14\% \\
  		 &     & 0.3 & 0.9 & 345 & 85.39\% & 309 & 81.08\% \\
  		 &     & 0.5 & 0.9 & 726 & 84.74\% & 653 & 80.51\% \\[0.1cm]
2:2:1  & 0.6 & 0.1 & 0.5 & 270 & 78.59\% & 283 & 80.36\% \\
  		 &		 & 0.1 & 0.7 & 115 & 79.96\% & 119 & 80.62\% \\
  		 &		 & 0.1 & 0.9 & 50  & 84.71\% & 49  & 80.71\% \\
  		 &		 & 0.3 & 0.7 & 290 & 80.73\% & 287 & 80.02\% \\
  		 &		 & 0.3 & 0.9 & 95  & 84.25\% & 89  & 80.82\% \\
  		 &		 & 0.5 & 0.9 & 213 & 86.06\% & 186 & 81.11\% \\[0.1cm]
  		 & 0.8 & 0.1 & 0.7 & 510 & 81.69\% & 492 & 80.15\% \\
  		 &     & 0.1 & 0.9 & 170 & 85.42\% & 156 & 81.99\% \\
  		 &     & 0.3 & 0.9 & 300 & 85.51\% & 269 & 81.09\% \\
  		 &     & 0.5 & 0.9 & 635 & 84.69\% & 575 & 80.88\% \\[0.1cm]
3:2:1  & 0.6 & 0.1 & 0.5 & 252 & 78.15\% & 268 & 80.49\% \\
  		 &		 & 0.1 & 0.7 & 108 & 80.54\% & 110 & 81.05\% \\
  		 &		 & 0.1 & 0.9 & 42  & 80.12\% & 45  & 83.09\% \\
  		 &		 & 0.3 & 0.7 & 276 & 80.97\% & 272 & 80.31\% \\
  		 &		 & 0.3 & 0.9 & 90  & 85.70\% & 83  & 81.07\% \\
  		 &		 & 0.5 & 0.9 & 204 & 87.31\% & 173 & 80.65\% \\[0.1cm]
  		 & 0.8 & 0.1 & 0.7 & 486 & 82.51\% & 458 & 80.17\% \\
  		 &     & 0.1 & 0.9 & 156 & 87.36\% & 135 & 81.75\% \\
  		 &     & 0.3 & 0.9 & 282 & 87.17\% & 241 & 81.21\% \\
  		 &     & 0.5 & 0.9 & 606 & 86.02\% & 520 & 80.30\% \\[0.1cm]
\hline
  		       
\end{tabular}
\end{table}  
}

\subsubsection{Optimal allocation}\label{sec:bin_opt}
For binary distributed endpoints and the hypothesis (\ref{eqn:hyp_bin}) the optimal allocation of samples is given by
\begin{eqnarray}\label{bin_opt2}
n_T^*:n_R^*:n_P^*\;=\;1:\Delta\:\sqrt{\frac{\pi_{0,R}(1-\pi_{0,R})}{\pi_{0,T}(1-\pi_{0,T})}}:|1-\Delta|\:\sqrt{\frac{\pi_{0,P}(1-\pi_{0,P})}{\pi_{0,T}(1-\pi_{0,T})}}
\end{eqnarray}
according to (\ref{opt_alloc}). For the commonly used alternative $\pi_{0,R}=\pi_{0,T}$ the allocation simplifies to
\begin{eqnarray}\label{bin_opt3}
n_T^*:n_R^*:n_P^*\;=\;1:\Delta\::|1-\Delta|\:\sqrt{\frac{\pi_{0,P}(1-\pi_{0,P})}{\pi_{0,T}(1-\pi_{0,T})}}\;.
\end{eqnarray}
In contrast to the case of normally distributed endpoints, where the optimal allocation is given by $1:\Delta\::|1-\Delta|$ (cf. Pigeot et al., 2003), the optimal allocation depends on the parameter of investigation. 

Kieser \& Friede (2007) derived the optimal allocation under the additional constraint that the test and reference group are balanced, $n_T^*=n_R^*$. Our result (\ref{bin_opt2}) shows that this restriction does not lead to an approximative optimal allocation, in general. Exemplary, Kieser \& Friede (2007) derive that the allocation $2.1:2.1:1$ would be optimal for $\pi_P=0.1$, $\pi_T=\pi_R=0.9$ and $\Delta=0.6$, whereas (\ref{bin_opt3}) yields an optimal allocation of $2.5:1.5:1$, giving more weight to the test group relative to the reference group and nearly the same to the placebo group. The allocation $2.5:1.5:1$ and the allocation $2.1:2.1:1$ result in a total required sample size of 79 and 89, respectively, when a power $1-\beta$ of 80\% is desired. Thus, our optimal allocation makes a further reduction of total sample size of about 12\% possible in this specific setting. The sample size reductions which are possible in other settings are illustrated in Figure \ref{fig:red2}, where the reduction for the optimal allocation instead of a balanced and a 2:2:1 allocation, respectively, is presented for $\pi_{0,P}=0.1$ and different values of $\Delta$, exemplary. For the 2:2:1 allocation we observe reductions between about 3\% and 10\%. For the balanced allocation there are reductions up to 20\% and more possible. Thus, the 2:2:1 allocation is more apporiate than the balanced allocation. However, it can be further improved by the optimal one (\ref{bin_opt3}).

\subsubsection{Planning a trial - applying the GSSP}

For binary distributed endpoints the weighted KL-divergence is given by
\begin{eqnarray}\label{eqn:KL-Bin}
K(\zeta^{(0)},\zeta,w)=\sum_{k=T,R,P} w_k \cdot \left(\pi_k^{(0)} \cdot \log \frac{\pi_k^{(0)}}{\pi_k} + (1-\pi_k^{(0)}) \cdot \log \frac{1-\pi_k^{(0)}}{1-\pi_k}\right)
\end{eqnarray}
with $\zeta=(\pi_T,\pi_R,\pi_P)$ and $\zeta^{(0)}=(\pi_T^{(0)},\pi_R^{(0)},\pi_P^{(0)})$. We restrict our investigations in the following to the commonly used alternative $\pi_T^{(0)}=\pi_R^{(0)}$. To restrict the minimization problem of the weighted KL-divergence to $H_{0,\pi_k}$ (\ref{eqn:hyp_bin}) we substitute $\pi_T=\Delta \pi_R + (1-\Delta) \pi_P$ in (\ref{eqn:KL-Bin}). We have minimized the KL-divergence (\ref{eqn:KL-Bin}) in $\pi_R$ and $\pi_P$ by the Newton-Raphson algorithm, confer Section \ref{sec:KL-practice}. Note, that this is a strictly convex optimization problem by Theorem 1 because 
$$
-E_{\pi_k^{(0)}}\left[\frac{\partial^2}{\partial^2\pi}\log f(\pi,X)\right]=\frac{\pi_k^{(0)}}{\pi^2}+\frac{1-\pi_k^{(0)}}{(1-\pi)^2}> 0
$$ 
for any $\pi$ and $\pi_k^{(0)}$. This guarantees the existence of a unique minimizer and geometric convergence of the Newton-Raphson algorithm. Based on the obtained results the limit $\sigma_{RML}^2$ of the restricted MLE's of the variance is computed and compared to the true variance $\sigma_0^2$, see Table \ref{table_bin}, columns 6 and 9. We used throughout Table \ref{table_bin} a choice of $\Delta=0.7$, exemplary.

We may use (\ref{samp_for_1}) and (\ref{eqn:samplesize}), respectively, to compute the total required sample sizes. The results are also displayed in Table \ref{table_bin} for a power $1-\beta$ of 0.7 and 0.8, respectively, for the optimal allocation, displayed in the columns 3-5 of Table \ref{table_bin}, and the commonly used 2:2:1 allocation for the purpose of illustrating the influence of allocation on the total required sample size. The sample size values in brackets are determined by (\ref{samp_for_1}), i.e. the RET is performed with unrestricted estimation of variance $\hat{\sigma}^2=\hat{\sigma}^2_{ML}$, and the values in front without brackets are determined by (\ref{eqn:samplesize}), i.e. the RET is performed with restricted estimation of variance $\hat{\sigma}^2=\hat{\sigma}^2_{RML}$. For large sample sizes the differences between both values are relatively small, whereas for small to moderate sample sizes ($n<200$) the differences are more pronounced. The amount of difference is driven by the difference between $\sigma_{RML}$ and $\sigma_0$, see again Table \ref{table_bin}, column 6 and 9.

It is important to note, that these results differ from those obtained by Kieser \& Friede (2007). This is due to the fact, that for the computation of  $\sigma_{RML}^2$ we have used the limit of the restricted MLE $\hat{\sigma}_{RML}^2$ instead of choosing an arbitrary parameter constellation on the boundary of $H_{0,\pi_k}$. We will see that the usage of the exact limit $\sigma_{RML}^2$ improves significantly the precision of the sample size formula (\ref{eqn:samplesize}). To this end, we have determined the required total sample size $n$ via (\ref{eqn:samplesize}) with usage of the exact limit $\sigma_{RML}^2$ to obtain a power of 80\% at level $\alpha=2.5\%$ (in order to be comparable with the results obtained by Kieser \& Friede (2007)) for different parameter settings and allocations and thereafter we have computed the resulting exact power (see Table \ref{table_bin2}). Note, that we always have rounded down the group sample sizes $n_k$, $k=T,R,P$. The results obtained by Kieser \& Friede (2007), who have not used the exact limit $\sigma_{RML}^2$, are displayed for comparison. Kieser \& Friede (2007) obtain an exact power that increases to 85\% or even to 87\% for some settings although $n>200$. Whereas the power decreases up to 78\% for other settings. In contrast, our method results in power values between 80\% and 82\% for all settings (with one exception for the case $w_T:w_R:w_P=3:2:1$, $\Delta=0.6$, $\pi_{0,P}=0.1$ and $\pi_{0,R}=0.9$ due to the small total sample size of 45). In summary, we find that our approximative formula yields very satisfactory results over a broad range of scenarios.

\subsection{Poisson endpoints: Treatment of epilepsy}\label{ex:pois2}

\begin{figure}[t]
\centering
\includegraphics[height=6cm, angle=-90]{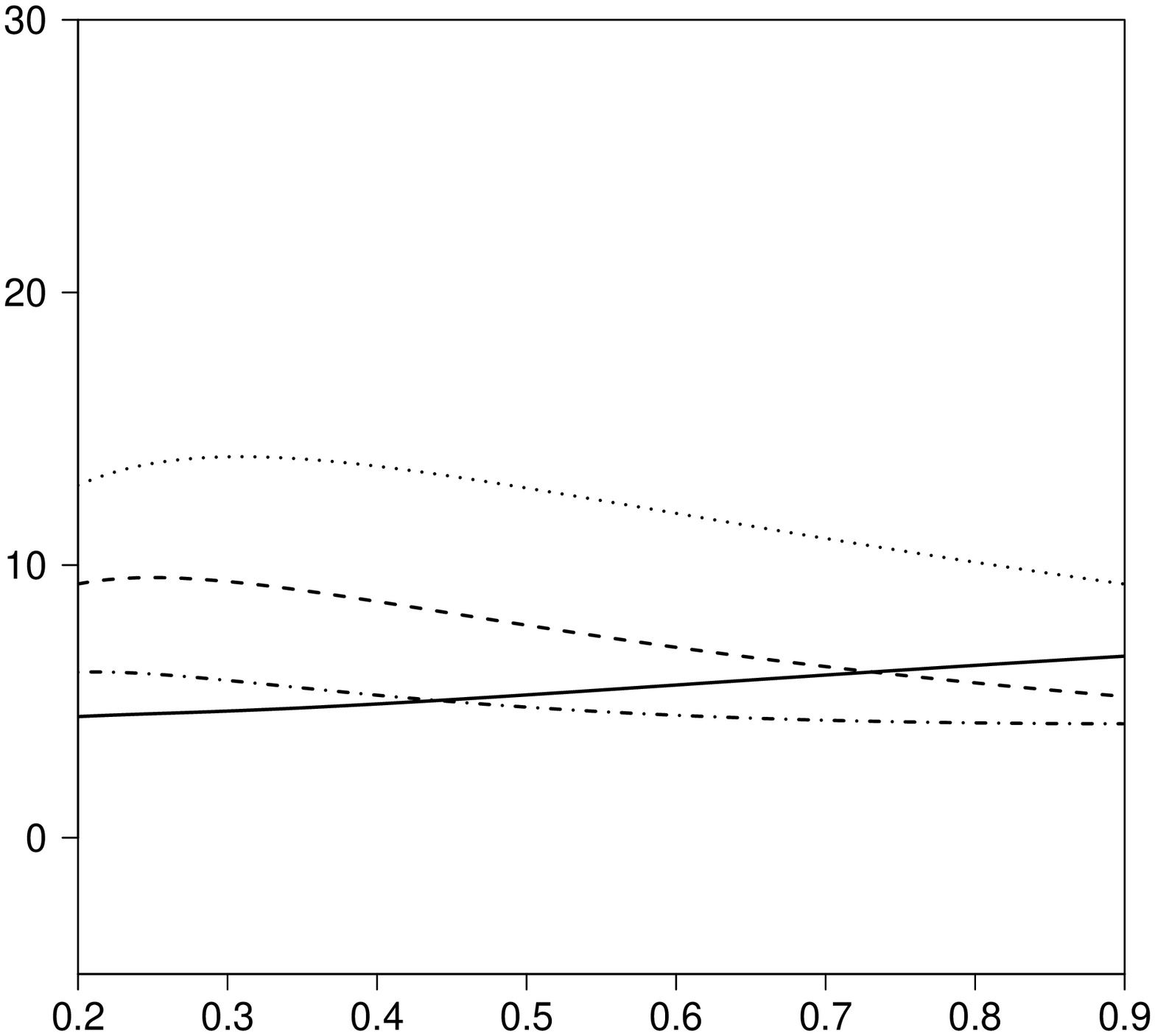}
\put(-90,-170){$\lambda_{0,R}/\lambda_{0,P}$}
\put(-180,-80){\%}
\hspace{1cm}
\includegraphics[height=6cm, angle=-90]{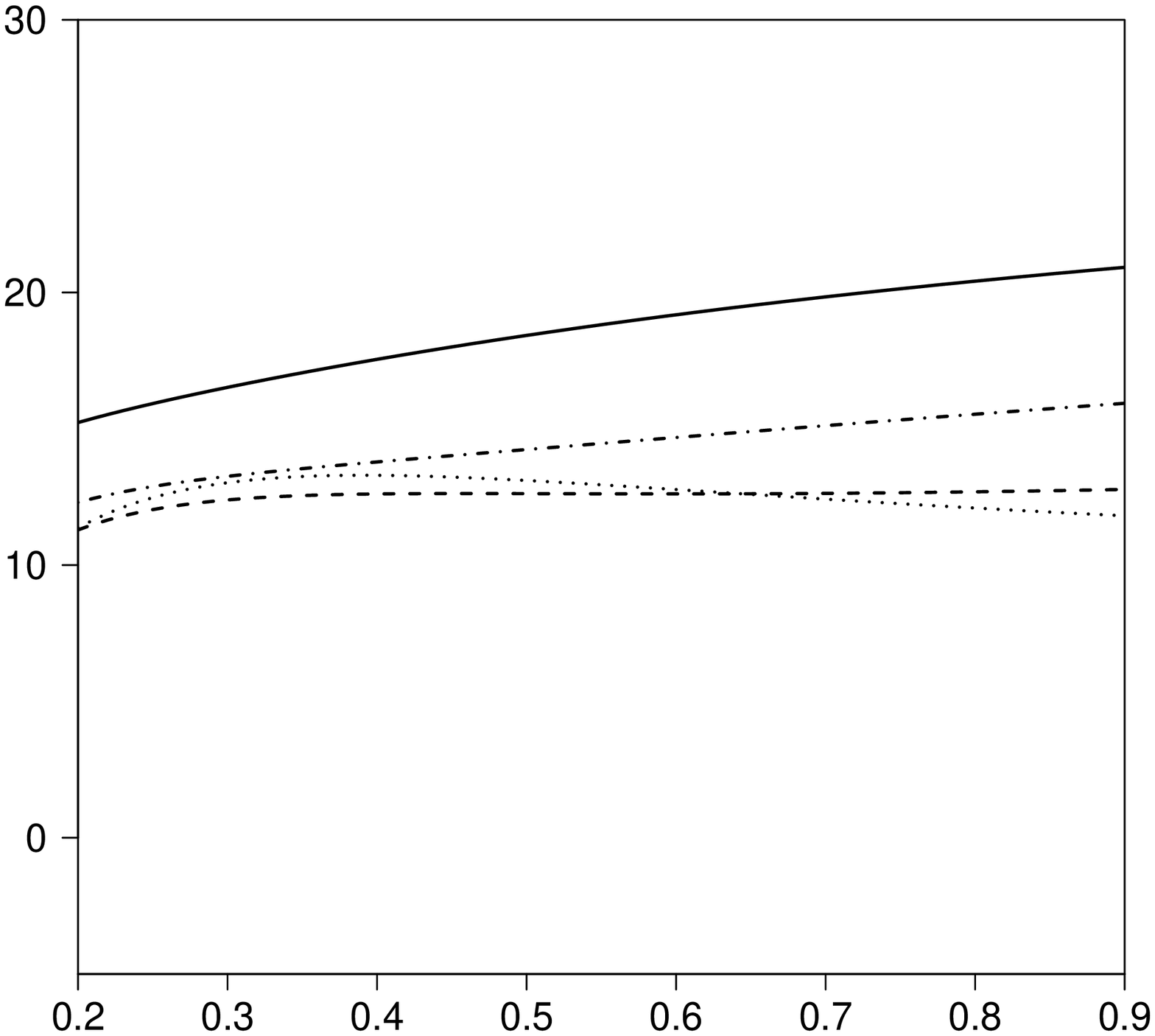}
\put(-90,-170){$\lambda_{0,R}/\lambda_{0,P}$}
\put(-180,-80){\%}

\caption{Example of Poisson distributed endpoints: \textbf{Sample size reduction in \%} when optimal allocation is used instead of the balanced allocation (right figure) and instead of the allocation 2:2:1 (left figure) for different values of $\Delta$, $\Delta=0.5$ (dotted line), $\Delta=0.6$ (dashed line), $\Delta=0.7$ (dotdash line), $\Delta=0.8$ (solid line). } \label{fig:red1}
\end{figure}

In this section, we revisit the example in the treatment of epilepsy introduced in Section \ref{sec:ex_pois}. 

\subsubsection{Performing the RET} 
The MLE $\hat{\lambda}_k$ is obtained by the mean value $n_k^{-1}\sum_{i=1}^{n_k}X_{ki}$, which is asymptotically normally distributed with variance $\sigma_k^2=\lambda_k$. The unrestricted MLE of the variance $\sigma^2$ is obtained by 
$$
\hat{\sigma}^2_{ML}=n\cdot\left(\frac{\hat{\lambda}_T}{n_T}+\Delta^2 \frac{\hat{\lambda}_R}{n_R}+ (1-\Delta)^2\frac{\hat{\lambda}_P}{n_P} \right)\:.
$$
Hence, we end up with the test statistic (see (\ref{eqn:teststatistic}))
\begin{eqnarray}\label{eqn:teststat_pois}
T=\frac{-\hat{\lambda}_T+\Delta\hat{\lambda}_R+(1-\Delta)\hat{\lambda}_P}{\sqrt{\frac{\hat{\lambda}_T}{n_T}+\Delta^2 \frac{\hat{\lambda}_R}{n_R}+ (1-\Delta)^2\frac{\hat{\lambda}_P}{n_P}}}
\end{eqnarray}
in order to test $H_{0,-\lambda_k}$ from (\ref{eqn:hyp_pois}), where $H_{0,-\lambda_k}$ is rejected if $T>z_{1-\alpha}$. The restricted version of the Wald-type test is observed by replacing the MLEs $\hat{\lambda}_k$ in the denominator by the to $H_{0,-\lambda_k}$ restricted ones. Again, we have computed the restricted MLEs numerically as for binary endpoints in the previous section.

The RET for the hypothesis (\ref{eqn:hyp_pois}) with $\Delta=0.5$ yields $T=1.328$ $(1.349)$ in (\ref{eqn:teststat_pois}) using the restricted (unrestricted) estimator for the variance estimation and corresponding p-values 9.21\% (8.86\%). Thus, we would not reject $H_{0,-\lambda_k}$ from (\ref{eqn:hyp_pois}) at level $\alpha=0.05$ and we could not claim that the test treatment is non-inferior to the reference one. 

\subsubsection{Optimal allocation} 

{\footnotesize
\begin{table}[h]
\caption{Optimal allocation of samples for the example of Poisson distributed endpoints }\label{table2}
\begin{tabular}{|c|rrr|rrr|rrr|}
\hline
 & \multicolumn{3}{c|}{$\Delta=0.5$} & \multicolumn{3}{c|}{$\Delta=0.7$} &\multicolumn{3}{c|}{$\Delta=0.8$}\\
$\frac{\lambda_{0,T}}{\lambda_{0,P}}=\frac{\lambda_{0,R}}{\lambda_{0,P}}$ & $w_T^*$ & $w_R^*$ & $w_P^*$ & $w_T^*$ & $w_R^*$ & $w_P^*$ & $w_T^*$ & $w_R^*$ & $w_P^*$ \\[0.1cm]
\hline\hline
0.9&0.49&0.25&0.26&0.50&0.35&0.16&0.50&0.40&0.10\\
0.8&0.49&0.24&0.27&0.49&0.34&0.16&0.49&0.40&0.11\\
0.7&0.48&0.24&0.28&0.49&0.34&0.17&0.49&0.39&0.12\\
0.6&0.47&0.23&0.30&0.48&0.34&0.19&0.49&0.39&0.13\\
0.5&0.45&0.23&0.32&0.47&0.33&0.20&0.48&0.38&0.14\\
0.3&0.41&0.21&0.38&0.44&0.31&0.24&0.46&0.37&0.17\\
0.2&0.38&0.19&0.43&0.42&0.30&0.28&0.44&0.36&0.30\\[0.1cm]
\hline
\end{tabular}
\end{table}  
}
For Poisson distributed endpoints and the hypothesis (\ref{eqn:hyp_pois})  the optimal allocation of samples is given by
\begin{eqnarray}\label{eqn:pois_optalloc}
n_T^*:n_R^*:n_P^*\;=\;1:\Delta\:\sqrt{\frac{\lambda_{0,R}}{\lambda_{0,T}}}:|1-\Delta|\:\sqrt{\frac{\lambda_{0,P}}{\lambda_{0,T}}}\;.
\end{eqnarray}
Table \ref{table2} presents the optimal allocation for the commonly used alternative $\lambda_{0,T}=\lambda_{0,R}$ for different choices of $\lambda_{0,T}/\lambda_{0,P}=\lambda_{0,R}/\lambda_{0,P}$ and $\Delta$. Note, that we may assume w.l.o.g $\lambda_{0,P}=1$ because multiplication of all parameters $\lambda_{0,k}$, $k=T,R,P$, by the same factor does not change the optimal allocation. This simplifies computation significantly. The sample size reductions which are possible are illustrated in Figure \ref{fig:red1} where the reduction for using the optimal allocation instead of a balanced and a 2:2:1 allocation, respectively, is presented for different values of $\Delta$. The results are quite similar to the ones for binary endpoints in the previous section.

{\footnotesize
\begin{table}[h]
\caption{\textbf{Example of Poisson distributed endpoints:} Limits of restricted MLE's, limit of variance estimator $\hat{\sigma}_{RML}$ and required samples size to obtain a power of 0.7 and 0.8, respectively, when the variance is estimated restrictedly to the null-hypothesis (unrestrictedly), a nominal significance level $\alpha=5\%$, for different parameter constellations and choices of $\Delta$ for the optimal sample allocation in (\ref{eqn:pois_optalloc}). } \label{table3}
\begin{tabular}{|lc|rrr|rr|r|rr|}
\hline & & & & & & & & &  \\[-0.2cm]
$\Delta$ & $\frac{\lambda_{0,T}}{\lambda_{0,P}}=\frac{\lambda_{0,R}}{\lambda_{0,P}}$ & $\frac{\lambda_{T,H_0}}{\lambda_{0,P}}$ &
$\frac{\lambda_{R,H_0}}{\lambda_{0,P}}$ & $\frac{\lambda_{P,H_0}}{\lambda_{0,P}}$ & $\frac{\sigma_{RML}}{\lambda_{0,P}}$ & $\frac{\sigma_0}{\lambda_{0,P}}$ & $\frac{\sigma_{RML}}{\sigma_0}$&$n_{0.7}\cdot \lambda_{0,P}$&$n_{0.8}\cdot \lambda_{0,P}$\\[0.2cm]
\hline\hline & & & & & & & & &  \\[-0.2cm]
0.5&0.7&0.78&0.64&0.92&1.763&1.755&1.005&649 (645)&852 (847)\\
   &0.5&0.64&0.41&0.87&1.594&1.561&1.021&190 (184)&248 (241)\\
   &0.3&0.51&0.21&0.81&1.426&1.322&1.079&76 (68)&98 (89)\\[0.2cm]
0.7&0.7&0.75&0.66&0.95&1.726&1.722&1.002&1729 (1724)&2270 (2265)\\
   &0.5&0.58&0.44&0.91&1.515&1.502&1.009&479 (472)&628 (620)\\
   &0.3&0.42&0.23&0.86&1.278&1.231&1.038&172 (162)&224 (213)\\[0.2cm]
0.8&0.7&0.73&0.67&0.97&1.707&1.706&1.001&3810 (3805)&5004 (4999)\\
   &0.5&0.55&0.46&0.94&1.479&1.473&1.004&1028 (1021)&1349 (1342)\\
   &0.3&0.38&0.25&0.90&1.210&1.186&1.020&348 (338)&456 (444)\\[0.1cm]
\hline
\end{tabular}
\end{table}  
}

\subsubsection{Planning a trial - applying the GSSP}\label{plan_pois}
For Poisson distributed endpoints the weighted KL-divergence is given by
\begin{eqnarray}\label{eqn:KL-Pois}
K(\zeta^{(0)},\zeta,w)=\sum_{k=T,R,P} w_k \cdot \left(\lambda_k - \lambda_k^{(0)}+\lambda_k^{(0)}\cdot\left(\log\lambda_k^{(0)}-\log\lambda_k\right)\right)\
\end{eqnarray}
with $\zeta=(\lambda_T,\lambda_R,\lambda_P)$ and $\zeta^{(0)}=(\lambda_T^{(0)},\lambda_R^{(0)},\lambda_P^{(0)})$. In the following we restrict our investigations to the commonly used alternative $\lambda_T^{(0)}=\lambda_R^{(0)}$. To restrict the minimization problem of the weighted KL-divergence to the boundary of $H_{0,-\lambda_k}$ (\ref{eqn:hyp_pois}) we substitute $\lambda_T=\Delta \lambda_R + (1-\Delta) \lambda_P$ in (\ref{eqn:KL-Pois}). For this situation, an explicit minimization of the KL-divergence is possible. To this end, we evaluate the derivatives of $K$ w.r.t. $\lambda_R$ and $\lambda_P$ at zero which is extremely cumbersome and yields a rather complex solution (see \ref{appendix:example}). The KL-divergence minimizer over $H_{0,-\lambda_k}$, denoted by $\lambda_{k,H_0}$, $k=T,R,P$, are displayed in Table \ref{table3} (columns 3-5) for different parameter constellations and choices of $\Delta$. Based on these results the limit $\sigma_{RML}^2$ of the restricted MLE's of the variance is computed (column 6) and compared to the true variance $\sigma_0^2$, see Table \ref{table3} columns 7 and 8. We presumed throughout Table \ref{table3} the usage of the optimal allocation from Table \ref{table2}. In addition, for all parameter constellations the required total samples size $n_{0.7},n_{0.8}$ to obtain a power of 0.7 and 0.8, respectively, are computed via (\ref{samp_for_1}) (values in brackets) and (\ref{eqn:samplesize}), respectively. Note, that  $n_{1-\beta}\cdot\lambda_{0,P}$ is displayed in Table \ref{table3} and thus the displayed values have to be divided by $\lambda_{0,P}$ to obtain the required total sample sizes.

\section{Software}\label{sec:software}

We provide the R source code of functions and documentation for planning and analyzing the RET for various endpoints as supplementary material (File: \textit{RET.Package.pdf}). This covers binary (Section \ref{ex:bin2}), Poisson (Section \ref{ex:pois2}), normally (Pigeot et al., 2003) and censored, exponentially distributed endpoints (Mielke et al., 2008). All provided functions have the following common structure:\\[-0.1cm]

{\footnotesize
\begin{table}[h]
\begin{tabular}{l@{\hspace{1cm}}l}
\hline \\[-0.2cm]
\textit{RET.xx.yy(\:)} & Performs the RET for given data\\[0.2cm]
\textit{RET.xx.yy.OptAlloc(\:)} & Computes the optimal sample allocation for the RET\\[0.2cm]
\textit{RET.xx.yy.Samplesize(\:)} & Determines the required sample sizes for the RET\\[0.2cm]
\hline
\end{tabular}
\end{table}
}
  
\noindent where 'xx' specifies the distribution of the endpoints and 'yy' the retention of effect hypothesis.

\section{Discussion}\label{sec:disc}

In this paper, we have presented a full analysis and planning of three-armed trials for general retention of effect hypotheses. The endpoint of interest may follow any (regular) parametric distribution family. As a major result, we have derived the asymptotically optimal allocation, see Equation (\ref{opt_alloc}), and sample size formulas for planning the trial (\ref{samp_for_1}) and (\ref{eqn:samplesize}) for restricted as well as unrestricted estimation of the variance. To this end, the crucial step was the determination of the exact limit $\sigma^2_{RML}$ of the restricted MLE of the variance $\sigma^2$, which was not investigated and incorporated in this context so far to our knowledge. As a consequence, note that for planning a non-inferiority trial it is important to decide in advance which estimation method will be performed as it affects the power and hence the total number of samples required. 

For binomial endpoints this improves on existing procedures. This includes the precision of the sample size formula  as well as the issue of optimal allocation. The optimal allocation reduces the total sample size by amounts up to 10\% (20\%) compared to the 2:2:1 (balanced) allocation. In addition, the methods of this paper are applied to Poisson endpoints, which were not investigated in the context of three-armed non-inferiority trials so far to our knowledge.

A problematic issue might be that the sample size planning and evaluation of a study presented in this paper is based on asymptotically considerations. Thus, for finite samples the optimal allocation could differ. In both examples investigated in this paper this is not the case, at least numerical studies show that the differences are irrelevantly small. However, differences could occur for example when the ratio $\sigma_{RML}/\sigma_0$ is far away from 1 and the signal to noise ratio $\eta{(0)}/\sigma_0$ is very small.

\appendix

\section{}

\subsection{Limit of the restricted MLE}\label{appendix:proof}

\noindent \textit{Assumption 1}: For $\zeta^{(0)}$ in the alternative $H_1$ and $n_k/n \rightarrow w_k \in ]0,1[$, $w=(w_T,w_R,w_P)$, 
the minimum $\zeta_{H_0}=(\theta_{T,H_0},\theta_{R,H_0},\theta_{P,H_0})=\arg\min_{\zeta\in H_0} K(\zeta^{(0)},\zeta,w)$ is well-defined.\\

\noindent\textit{Assumption 2}: For any sequence $\zeta^{(n)}=(\theta_T^{(n)},\theta_R^{(n)},\theta_P^{(n)})$ in $H_0$ with $\lim_{n\rightarrow \infty} \zeta^{(n)}$ in $\overline{\Theta}^3\setminus \Theta^3$ or with $\lim_{n\rightarrow \infty} \parallel \zeta^{(n)} \parallel = \infty$
$$
\lim_{n\rightarrow \infty} \prod_{k=T,R,P} f(\theta_k^{(n)},x_k)\;=\;0
$$
holds $P^{\zeta^{(0)}}$ almost everywhere. 

\smallskip
The next theorem shows that the restricted MLE converges to the minimizer of the sample size weighted Kullback-Leibler-divergence (KL-divergence) with respect to the true parameter, denoted by $\theta_{k,H_0}$, $k=T,R,P$.

\smallskip 

\noindent \textbf{Theorem 2}: \textit{Let $\hat{\zeta}_n^{H_0}$ denote the MLE restricted to $H_0$. Then under the Assumptions 1 and 2
$$
\hat{\zeta}_n^{H_0} \stackrel{a.s.}{\longrightarrow} \zeta_{H_0}\;.
$$}

\noindent\textit{Proof.} Let 
$$
Q_n(\zeta) = - \sum_{k=T,R,P} \frac{1}{n}\sum_{i=1}^{n_k} \log f(X_{ki},\theta_k)
$$
and
$$
Q(\zeta) = -\sum_{k=T,R,P} w_k\cdot E_{\theta_k^{(0)}}[\log f (X_{k1},\theta_k)].
$$
Note, that by definition
$$
K(\zeta^{(0)},\zeta,w)=Q(\zeta)-Q(\zeta^{(0)})
$$
holds and consequently $\zeta_{H_0}=\arg\min_{\zeta\in H_0} K(\zeta^{(0)},\zeta,w)$ is also the well-defined minimizer of $Q(\zeta)$ in $H_0$. 

Assumption 2 ensures that the MLE is asymptotically almost surely located in a compact set, i.e. there exists compact subset $\tilde{H}_0$ such that 
$$
\lim_{n\rightarrow\infty} \hat{\zeta}_n^{H_0} = \lim_{n\rightarrow\infty} \hat{\zeta}_n^{\tilde{H}_0}\hspace{1cm}a.s.
$$
A proof for $\lim_{n\rightarrow \infty} \parallel \zeta^{(n)} \parallel = \infty$ can be found in Wald (1949). However, for $\lim_{n\rightarrow \infty} \zeta^{(n)}$ in $\overline{\Theta}^3\setminus \Theta^3$ this can be proved analogously. Hence, we assume w.l.o.g. that $H_0$ is compact.
Therefore, the convergence
$$
Q_n(\zeta) \stackrel{a.s.}{\longrightarrow} Q(\zeta)
$$
is uniformly in $H_0$ (see Jennrich, 1969,Theorem 2) and we can apply Lemma 2.2 from White (1980), which yields that $\hat{\zeta}_n^{H_0}=\arg\min_{\zeta\in H_0}Q_n(\zeta)$ converges almost surely to the well-defined minimum $\zeta_{H_0}$ of $Q(\zeta)$ in $H_0$. 

\vspace{-0.3cm}\hspace{13cm}$\qed$

\subsection{Proof of Theorem 1}\label{proof:thm1}

The condition $-E_{\theta_k^{(0)}}[\frac{\partial^2}{\partial^2\theta}\log f(\theta,X)]$ ensures that the KL-divergence $K(\theta_k^{(0)},\theta)$ is a convex function in $\theta$ for $\theta_k^{(0)}$, $k=T,R,P$. Thus, the weighted KL-divergence 
$$
K(\zeta^{(0)},(\theta_T,\theta_R,\theta_P),c)= \sum_{k=T,R,P} c_k \cdot K(\theta_k^{(0)},\theta_k) 
$$
is convex in the arguments $\theta_k$, $k=T,R,P$. Let us denote 
$$
g(\theta_R,\theta_P)=h^{-1}(\Delta h(\theta_R)+(1-\Delta)h(\theta_P)),
$$
which is an affine transformation in both arguments by assumption. Hence, 
$$
K(\theta_T^{(0)},g(\theta_R,\theta_P))
$$
is a convex function in $(\theta_R,\theta_P)$. Therefore, the weighted KL-divergence with restriction to the boundary of the null hypothesis $H_{0,h(\theta_k)}$ represented by
$$
K(\zeta^{(0)},(g(\theta_R,\theta_P),\theta_R,\theta_P),c)=c_T \cdot K(\theta_T^{(0)},g(\theta_R,\theta_P))+ c_R \cdot K(\theta_R^{(0)},\theta_R) + c_P \cdot K(\theta_P^{(0)},\theta_P)
$$
is a linear combination of convex function and therewith convex in $(\theta_R,\theta_P)$, again.

\vspace{0.1cm}\hspace{13cm}$\qed$ 

\subsection{Minimization of $\sigma^2_0$ as a function of sample allocation}\label{app:min_var}
As the sample allocation has to fulfill $w_T+w_R+w_P=1$ we substitute $w_P=1-w_T-w_R$ in (\ref{var_0}) and obtain
$$
\sigma_0^2 = \frac{\sigma_{0,T}^2}{w_T}+\frac{\Delta^2 \cdot\sigma_{0,R}^2}{w_R}+\frac{(1-\Delta)^2\cdot\sigma_{0,P}^2}{1-w_T-w_R}\:.
$$ 
Note, that $\sigma_0^2$ is convex function in $(w_T,w_R)$. Evaluating the derivatives of $\sigma_0^2$ w.r.t. $w_T$ and $w_R$ at zero yields
\begin{eqnarray*}
\frac{\partial}{\partial w_T}\:\sigma_{0}^2 &=& \frac{(1-\Delta)^2 \cdot \sigma_{0,P}^2}{(1-w_R-w_T)^2}-\frac{\sigma_{0,T}^2}{w_T^2} = 0 \\
\frac{\partial}{\partial w_R}\:\sigma_{0}^2 &=& \frac{(1-\Delta)^2 \cdot \sigma_{0,P}^2}{(1-w_R-w_T)^2}-\frac{\Delta^2 \cdot \sigma_{0,T}^2}{w_T^2} = 0 .
\end{eqnarray*}
Solving the equations for $w_T$ and $w_R$ yields the minimizer
\begin{eqnarray*}
 w_T^* &=& \frac{\sigma_{0,T}}{\sigma_{0,T}+\Delta\cdot\sigma_{0,R}+|1-\Delta|\cdot\sigma_{0,P}}\\
 w_R^* &=& \frac{\Delta \cdot \sigma_{0,R}}{\sigma_{0,T}+\Delta\cdot\sigma_{0,R}+|1-\Delta|\cdot\sigma_{0,P}}
\end{eqnarray*}
and therewith
\begin{eqnarray*}
 w_P^* &=& \frac{|1-\Delta|\cdot \sigma_{0,P}}{\sigma_{0,T}+\Delta\cdot\sigma_{0,R}+|1-\Delta|\cdot\sigma_{0,P}}\:.
\end{eqnarray*}
Thus, the optimal allocation in terms of minimizing the variance $\sigma^2_0$ is given by
\begin{eqnarray*}
n_T^*:n_R^*:n_P^*=w_T^*:w_R^*:w_P^*\;=\;1:\Delta\:\frac{\sigma_{0,R}}{\sigma_{0,T}}:|1-\Delta|\:\frac{\sigma_{0,P}}{\sigma_{0,T}}\:.
\end{eqnarray*}

\vspace{0.1cm}\hspace{13cm}$\qed$ 

\subsection{Comparison of the variance $\sigma_0^2$ for different allocations}\label{sec:comp_var}
\textbf{Theorem 3:} \textit{ If $\theta_R^{(0)}=\theta_T^{(0)}$ and  $\sigma_{0,P}^2/\sigma_{0,T}^2<2.12$ then the allocation $1:\Delta:(1-\Delta)$ results in a smaller variance $\sigma_0^2$ (\ref{var_0}) (and hence larger asymptotic power) than the allocation 2:2:1 for any $0\leq\Delta\leq1$.}
\vspace{0.2cm}

\noindent \textit{Proof.} Substituting the allocation 2:2:1 and $1:\Delta:(1-\Delta)$, respectively, and $\theta_R^{(0)}=\theta_T^{(0)}$ in the variance $\sigma_0^2$ from (\ref{var_0}) yields
$$
\sigma_{2:2:1}^2= \frac{5+5\Delta^2}{2}\: \sigma_{0,T}^2\;+\; 5(1-\Delta)^2\:\sigma_{0,P}
$$
and
$$
\sigma_{1:\Delta:1-\Delta}^2=2(1+\Delta)\: \sigma_{0,T}^2\;+\; 2(1-\Delta)\:\sigma_{0,P}^2.
$$
Thus, we obtain with $r:=\sigma_{0,P}^2/\sigma_{0,T}^2>0$
$$
g(\Delta,r):=\frac{\sigma_{2:2:1}^2-\sigma_{1:\Delta:1-\Delta}^2}{\sigma_{0,T}^2}= \left(2.5+5\cdot r\right) \Delta^2 +  \left(-2-8\cdot r\right) \Delta + \left(0.5+3\cdot r\right),
$$
which is as a quadratic function in $\Delta$ with minimum 
$$
(a(r),b(r))=\left(\frac{2+8\cdot r}{5+10\cdot r} \:,\: \frac{-4\cdot(r-2.11803)(r+0.118034)}{10+20\cdot r}\right),
$$
where $0<a(r)<1$ and $b(r)>0$ for $r<2.11803\approx 2.12$. Thus, we obtain for $r=\sigma_{0,P}^2/\sigma_{0,T}^2<2.12$ that $g(\Delta,r)>0$, which implies $\sigma_{2:2:1}^2>\sigma_{1:\Delta:1-\Delta}^2$, for any $0\leq\Delta\leq1$.

\vspace{0.1cm}\hspace{13cm}$\qed$ 

\noindent\textbf{Theorem 4:} \textit{ If $\theta_R^{(0)}=\theta_T^{(0)}$ and  $\sigma_{0,P}^2/\sigma_{0,T}^2<2.73$ then the allocation $1:\Delta:(1-\Delta)$ results in a smaller variance $\sigma_0^2$ (\ref{var_0}) (and hence larger asymptotic power) than the balanced allocation 1:1:1 for any $0\leq\Delta\leq1$.}
\vspace{0.2cm}

\noindent \textit{Proof.} Substituting the allocation 1:1:1 and $1:\Delta:(1-\Delta)$, respectively, and $\theta_R^{(0)}=\theta_T^{(0)}$ in the variance $\sigma_0^2$ from (\ref{var_0}) yields
$$
\sigma_{1:1:1}^2= (3+3\Delta^2)\: \sigma_{0,T}^2\;+\; 3(1-\Delta)^2\:\sigma_{0,P}
$$
and
$$
\sigma_{1:\Delta:1-\Delta}^2=2(1+\Delta)\: \sigma_{0,T}^2\;+\; 2(1-\Delta)\:\sigma_{0,P}^2.
$$
Thus, we obtain with $r:=\sigma_{0,P}^2/\sigma_{0,T}^2>0$
$$
g(\Delta,r):=\frac{\sigma_{1:1:1}^2-\sigma_{1:\Delta:1-\Delta}^2}{\sigma_{0,T}^2}= \left(3+3\cdot r\right) \Delta^2 +  \left(-2-4\cdot r\right) \Delta + \left(1+r\right),
$$
which is as a quadratic function in $\Delta$ with minimum
$$
(a(r),b(r))=\left(\frac{2+4\cdot r}{6+6\cdot r} \:,\: \frac{-(r-1+\sqrt{3})(r-1-\sqrt{3})}{3+3\cdot r}\right),
$$
where $0<a(r)<1$ and $b(r)>0$ for $r<1+\sqrt{3}\approx 2.73$. Thus, we obtain for $r=\sigma_{0,P}^2/\sigma_{0,T}^2<2.73$ that $g(\Delta,r)>0$, which implies $\sigma_{1:1:1}^2>\sigma_{1:\Delta:1-\Delta}^2$, for any $0\leq\Delta\leq1$.

\vspace{0.1cm}\hspace{13cm}$\qed$ 
  
\subsection{Poisson example: weighted KL-divergence minimizer}\label{appendix:example}
An analytical solution to the minimization of the KL-divergence $K(\zeta^{(0)},\zeta,w)$ for Poisson endpoints in Section \ref{plan_pois} can be obtained by evaluating the derivatives of KL-divergence (\ref{eqn:KL-Pois}) w.r.t. $\lambda_R$ and $\lambda_P$ at zero after substituting $\lambda_T=\Delta\lambda_R+ (1-\Delta)\lambda_P$, which yields
\begin{eqnarray*}
\lambda_{R,H_0} &=& \left [ \Delta^2 (-1+w_T)w_T\lambda_{0,P}-\Delta(-1+w_T)w_T(\lambda_{0,P}-\lambda_{0,T})+w_R^2((-1+\Delta)\lambda_{0,P}+(2-\Delta)\lambda_{0,T})\right.\\ 
& +& \left. w_R((-1+\Delta)(-1+w_T+\Delta w_T)\lambda_{0,P}+(-\Delta+w_T+2\Delta w_T-\Delta^2 w_T)\lambda_{0,T}) - \mathbf{S}\right ]\\
& /& \left(2 (w_R+\Delta (-1+w_T)) (w_R+\Delta w_T)\right)\\[0.2cm]
\lambda_{P,H_0} &=& \left [ w_R^2 \lambda_{0,P}+\Delta^2 w_T ((-1+w_R+w_T) \lambda_{0,P}-w_R  \lambda_{0,T})+w_R ((-1+w_T)\lambda_{0,P}w_T \lambda_{0,T})\right.\\
&+& \left. \Delta ((2+w_R^2-3 w_T+w_T^2+w_R (-3+2 w_T)) \lambda_{0,P}+(w_R-w_R^2+w_T-w_T^2) \lambda_{0,T}) - \mathbf{S} \right ]\\
&/& \left(2 ((-1+w_R) w_R+\Delta^2 (-1+w_T) w_T+\Delta (1-w_T+w_R (-1+2 w_T)))\right)\\[0.2cm]
\lambda_{T,H_0} &=& \Delta \lambda_{R,H_0} + (1-\Delta)\lambda_{P,H_0}
\end{eqnarray*}
with
\begin{eqnarray*}
\mathbf{S} &=&\left\{ -4 \Delta (-1+w_R+w_T) ((-1+w_R) w_R+\Delta^2 (-1+w_T) w_T+\Delta (1-w_T+w_R (-1+2 w_T))) \lambda_{0,P} \right.\\
&\cdot&  ((-1+w_R+w_T) \lambda_{0,P}-(w_R+w_T) \lambda_{0,T})+(\Delta^2 w_T ((-1+w_R+w_T) \lambda_{0,P}-w_R \lambda_{0,T}) \\
& +& w_R ((-1+w_R+w_T) \lambda_{0,P}-w_T \lambda_{0,T})+\Delta ((2-3 w_R+w_R^2-3 w_T+2 w_R w_T+w_T^2) \lambda_{0,P} \\
&+& \left. (w_R-w_R^2+w_T-w_T^2) \lambda_{0,T}))^2 \right\}^{1/2} .
\end{eqnarray*}

\section*{Acknowledgements}

M. Mielke acknowledges support of the Georg Lichtenberg program 'Applied Statistics \& Empirical Methods' and A. Munk support of DFG FOR916. 

\section*{References}

\begin{description}

\item {\small R.B. D'Agostino (ed.) (2003). Special issue: Non-inferiority trials: Advances in concepts and methodology. \textit{Statistics in
Medicine}, 22, 165--336.}

\item {\small I.S.F. Chan (1998). Exact tests of equivalence and efficacy with a non-zero lower bound for comparative studies. \textit{Statistics in Medicine}, 17, 1403--1413.} 

\item {\small H. Dette, M. Trampisch \& L.A. Hothorn (2009). Robust designs in non-inferiority three arm clinical trials with presence of heteroscedasticity. \textit{Statistics in Biopharmaceutical Research}, 1, 268--278.} 

\item {\small C.P. Farrington \& G. Manning (1990). Test statistics and sample size formulae for comparative binomial trials with null hypothesis of non-zero risk difference or non-unity relative risk. \textit{Statistics in Medicine}, 9, 1447--1454.} 

\item {\small D.J. Goldstein, Y. Lu, M.J. Detke, C. Wiltse, C. Mallinckrodt, M.A. Demitrack (2004). Duloxetine in the
treatment of depression. A double-blind placebo-controlled comparison with paroxetine. \textit{Journal of Clinical
Psychopharmacology}, 24, 389–-398.} 

\item {\small H.S. Gurm, J.S. Yadav, P. Fayad, B.T. Katzen, G.J. Mishkel, T.K. Bajwa, G. Ansel, N.E. Strickman, H. Wang, S.A. Cohen, J.M. Massaro \& D.E. Cutlip (2008). Long-term results of Carotid Stenting versus Endarterectomy in high-risk patients.
\textit{New England Journal of Medicine}, 358,1572--1579.}

\item {\small M. Hasler, R. Vonk \& L.A. Hothorn (2008). Assessing non-inferiority of a new treatment in a three-arm trial in the presence of heteroscedasticity. \textit{Statistics in Medicine}, 27, 490--503.} 

\item {\small V. Hasselblad \& D.F. Kong (2001). Statistical methods for comparison to placebo in active-control trials. \textit{Drug Information Journal}, 35, 435--449.}

\item {\small W.W. Hauck \& S. Anderson (1999). Some issues in the design and analysis of equivalence trials. \textit{Drug Information Journal}, 33, 109--118.}

\item {\small D. Hauschke \& I. Pigeot (2005). Establishing efficacy of a new experimental treatment in the "Gold Standard" design. \textit{Biometrical Journal}, 47, 782--786.}

\item {\small E.B. Holgrem (1999). Establishing equivalence by showing that a specified percentage of the effect of the active control over placebo is maintained. \textit{Journal
of Biopharmaceutical Statistics}, 9, 651--659.}

\item {\small H.M.J. Hung, S.J. Wang \& R. O'Neill (2009). Challenges and regulatory experiences with non-inferiority trial design without placebo arm. \textit{Biometrical Journal}, 51, 324-334.}

\item {\small Hypericum Depression Trial Study Group (2004). Effect of hypericum perforatum (St John's Wort) in major depressive disorder. \textit{Journal of the American Medical Association}, 287, 1807--1814.}

\item {\small R. Jennrich (1969). Asymptotic properties of non-linear least squares estimators. \textit{Ann. Math. Statist.}, 40, 633--643.} 

\item {\small B. Jones, P. Jarvis, J.A. Lewis, \& A.F. Ebbutt (1996). Trials to assess equivalence: the
importance of rigorous methods. \textit{British Medical Journal}, 313, 36-–39.}

\item {\small M. Kieser \& T. Friede (2007). Planning and analysis of three-arm non-inferiority trials with binary endpoints. \textit{Statistics in Medicine}, 26, 253--273.} 

\item {\small A. Koch \& J. R\"ohmel (2004). Hypothesis testing in the ``gold standard'' design for proving
the efficacy of an experimental treatment relative to placebo and a reference. \textit{J. Biopharm. Stat.}, 14, 315–-325.}

\item {\small A. Koch (2005). Discussion on ``Establishing efficacy of a new experimental treatment in the gold standard design''. \textit{Biometrical Journal}, 47, 792--793.}

\item {\small G.G. Koch \& C.M. Tangen (1999). Nonparametric analysis of covariance and its role in noninferiority clinical trials. \textit{Drug Information Journal}, 33, 1145--1159.}

\item {\small S. Lange \& G. Freitag (2005). Choice of delta: requirements and reality - results of a systematic review. \textit{Biometrical Journal}, 47, 12--27.}

\item {\small M.W.J. Layard \& J.N. Arvesen (1978). Analysis of Poisson data in crossover experimental designs. \textit{Biometrics},34, 421--428.} 

\item {\small M.\ Mielke, A. \ Munk \& A. \ Schacht  (2008). The assessment of non-inferiority in a gold standard design with censored, exponentially distributed endpoints. \textit{Statistics in Medicine},
27, 5093–-511.}

\item {\small R. Mohanraj \& M.J. Brodie (2003). Measuring the efficacy of antiepileptic drugs. \textit{Seizure}, 12, 413--443.} 

\item {\small A. Munk \& H.-J. Trampisch (eds.) (2005). Special issue: Therapeutic equivalence - clinical issues and statistical methodology in noninferiority trials. \textit{Biometrical Journal}, 47, 1--108.}

\item {\small A.A. Nierenberg \& E.C. Wright (1999). Evolution of remission as the new standard in the
treatment of depression. \textit{Journal of Clinical Psychiatry}, 60 (suppl. 22),7–11.}

\item {\small I. Pigeot, J. Sch\"afer, J. R\"ohmel \& D. Hauschke  (2003). Assessing non-inferiority of a new treatment in a three-arm clinical trial including placebo. \textit{Statistics in Medicine},
22, 883–-899.}

\item {\small J. R\"ohmel (1998). Therapeutic equivalence investigations: statistical considerations. \textit{Statistics
in Medicine}, 17, 1703–-1714.}

\item {\small J. R\"ohmel \& U. Mansmann (1999). Unconditional non-asymptotic one-sided test for independent binomial proportions when the interest lies in showing non-inferiority and/or superiority. \textit{Biometrical Journal}, 41, 149-170.}

\item {\small M. Rothmann et al. (2003). Design and analysis of non-inferiority mortality trials in oncology. \textit{Statistics in Medicine}, 22, 239--264.} 

\item {\small J.W.A.S. Sander, P.N. Patsalos, J.R. Oxley, M.J. Hamilton \& W.C. Yuen (1990). A randomized double-blind placebo-controlled add-on trial of lamotrigine in patients with severe epilepsy. \textit{Epilepsy Res.}, 6, 221-226.} 

\item {\small T. Schwartz \& J. Denne (2006). A two-stage sample size recalculation procedure for placebo- and active-controlled non-inferiority trials. \textit{Statistics in Medicine}, 45, 3396--3406.} 

\item {\small G. Skipka, A. Munk \& G. Freitag (2004). Unconditional exact tests for the difference of binomial probabilities - contrasted and compared. \textit{Computational Statistics \& Data Analysis}, 47, 757--773.} 

\item {\small M.-L. Tang \& N.-S. Tang (2004). Tests of noninferiority via rate difference for three-arm clinical trials with placebo.
   \textit{Journal of Biopharmaceutical Statistics}, 14, 337--347.}

\item {\small N.-S. Tang, M-L. Tang \& S.-F. Wang  (2007). Sample size determination for matched-pair equivalence
trials using rate ratio. \textit{Biostatistics}, 8, 625--631.}

\item {\small R. Temple \& S. Ellenberg (2000). Placebo-controlled trials and active-control trials in the
evaluation of new treatments. part 1: Ethical and scientific issues. \textit{Annals of Internal
Medicine}, 133, 455–-463.}

\item {\small A.W. van der Vaart (1998). {\sl Asymptotic Statistics.} Cambridge University Press.} 

\item {\small A. Wald (1949). Note on consistency of the maximum likelihood estimate. \textit{Ann. Math. Statist.}, 20, 595--601.} 

\item {\small H. White (1980). Nonlinear regression on cross-section data. \textit{Econometrica}, 48, 721--746.} 

\item {\small N. Yakhno, A. Guekht, A. Skoromets, N. Spirin, E. Strachunskaya, A. Ternavsky, K.J. Olsen  \& P.L. Moller  (2006). Lornoxicam quick-release compared with diclofenac potassium: subjects and methods. \textit{Clinical Drug Investigation}, 26, 267--277.} 

\end{description}

\end{document}